\documentclass[paper]{JHEP}
\usepackage{amsmath}
\usepackage{epsfig}

\def\al{\alpha}
\def\as{\alpha_{\mbox{\scriptsize s}}}
\def\aef{\alpha_{\mbox{\scriptsize eff}}}

\def\ee{e^+e^-}
\def\LQCD{\Lambda_{\mbox{\scriptsize QCD}}}
\def\MSbar{\overline{\mbox{MS}}}
\def\MSbar{\overline{\mbox{\scriptsize MS}}}
\def\GeV{\mathop{\rm Ge\!V}}
\def\xB{x_{\mbox{\scriptsize B}}}
\def\bq{\bar q}

\def\can{\tau}

\def\be{\beta}

\def\eps{\epsilon}
\def\de{\delta}

\def\om{\omega}
\def\gam{\gamma}

\def\ics{x}
\def\isp{\xi}

\def\out{\mbox{\scriptsize out}}
\def\mat{\mbox{\scriptsize mat}}

\def\cO#1{{\cal{O}}\left(#1\right)}
\def\half{\mbox{\small $\frac{1}{2}$}}

\def\PT{\mbox{\scriptsize PT}}
\def\NP{\mbox{\scriptsize NP}}

\def\conf{\delta}
\def\cp{\lambda^{\NP}}

\def\bnu{\bar{\nu}}
\def\Ko{K_{\out}}
\def\bKo{\bar{K}_{\out}}

\def\ka{\kappa}
\def\vka{\vec{\ka}}

\def\cF{{\cal{F}}}
\def\cE{{\cal{E}}}

\def\cM{{\cal{M}}}

\def\cR{{\cal{R}}}
\def\cP{{\cal {P}}}
\def\cI{{\cal {I}}}
\def\cC{{\cal {C}}}
\def\cS{{\cal {S}}}
\def\cA{{\cal{A}}}
 \newskip\humongous \humongous=0pt plus 1000pt minus 1000pt
   \newif\ifdtup

\def\ga{\mathrel{\mathpalette\fun >}}
\def\fun#1#2{\lower3.6pt\vbox{\baselineskip0pt\lineskip.9pt
  \ialign{$\mathsurround=0pt#1\hfil##\hfil$\crcr#2\crcr\sim\crcr}}}
\title{Out-of-plane QCD radiation in DIS \\ with high \boldmath{$p_t$}
  jets\footnote{Research supported in part by the EU Fourth Framework
    Programme, `Training and Mobility of Researchers', Network
    `Quantum Chromodynamics and the Deep Structure of Elementary
    Particles', contract FMRX-CT98-0194 (DG12 - MIHT).}}

\author{A.~Banfi, G.~Marchesini, G.~Smye \\
  Dipartimento di Fisica, Universit{\`a} di Milano--Bicocca and INFN,
  Sezione di Milano, Italy} 

\author{ G.~Zanderighi \\
    Department of Physics, University of Durham, Durham DH1 3LE,
    England}

\abstract{We present a QCD analysis of the cumulative
  out-of-event-plane momentum distribution in DIS process with
  emission of high $p_t$ jets. We derive the all-order resummed result
  to next-to-leading accuracy and estimate the leading power
  correction. We aim at the same level of accuracy which, in $\ee$
  annihilation, seems to be sufficient for making predictions.  As is
  typical of multi-jet observables, the distribution depends on the
  geometry of the event and the underlying colour structure. This
  result should provide a powerful method to study QCD dynamics, in
  particular to constrain the parton distribution functions, to
  measure the running coupling and to search for genuine
  non-perturbative effects.}


\preprint{
     Bicocca--FT--01/22\\
     IPPP/01/55\\
     DCPT/01/110\\
     hep-ph/0111157}

\begin{document}

\section{Introduction}
The success of the QCD description of $\ee$ event-shape variables
makes these observables 
useful tools to study features of QCD radiation~\cite{PTstandard}, to
measure the running coupling \cite{Exp-running} and to search for
genuine non-perturbative effects \cite{NPstandard,DMW,Exp-shape}.

The standard QCD analysis of event shapes involves resummations of all
perturbative (PT) terms which are double (DL) and single (SL)
logarithmically enhanced and matching with exact fixed order PT
results.  In addition, to make quantitative predictions, one needs to
consider also non-perturbative (NP) power corrections.  These
standards have been already achieved for a number of $2$-jet event
shapes ($1\!-\!T, C, M^2/Q^2, B$ in $\ee$ annihilation
\cite{PTstandard,Milan,broad} and $1\!-\!T,B$ in DIS
\cite{TDIS,MilanDIS}), i.e. for observables which vanish in the limit 
of two narrow jets.

Only recently has the attention moved to three-jet shapes in $\ee$
annihilation (thrust minor $T_m$ \cite{Tmin} and $D$-parameter
\cite{Dpar}).  These three-jet observables exhibit a rich 
geometry dependent structure due to the fact that they are sensitive
to large angle soft emission (intra-jet radiation).
These results have been extended to jet production in hadron
collisions \cite{hh}.  The main difference between processes with or
without hadrons in the initial state is that in the first case
jet-shape distributions are not collinear and infrared safe (CIS)
quantities, but are finite only after factorizing collinear singular
contributions from initial state radiation (giving rise to incoming
parton distributions at the appropriate hard scale).

In this paper we consider the DIS proton-electron process
\begin{equation}
  \label{eq:ep}
e + p\to e \>+\>\mbox{jets}\>+\> \ldots
\end{equation}
in which we select events with high $p_t$ jets 
with $p_t\sim Q$.  The dots represent the initial state jet and
intra-jet hadrons.  This process involves (at least) three jets: two
large $p_t$ jets (generated by two hard partons in the final state
recoiling one against the other) and the initial state jet (generated
by the incoming parton). For $p_t\sim Q$, the exchanged boson of
momentum $q$, with $Q^2=-q^2$, can be treated as elementary.

The observable we study is 
\begin{equation}
  \label{eq:Kout}
  \Ko={\sum_h}'\,|{p_{h}^{\out}}|\>.
\end{equation}
To avoid measurements in the beam region, the sum indicated by
$\sum_h'$ extends over all hadrons not in the beam direction.
Here $p_{h}^{\out}$ is the out-of-plane momentum of the hadron $h$
with the {\it event plane} defined as the plane formed by the proton
momentum $\vec{P}$ in the Breit frame and the unit vector $\vec{n}$
which enters the definition of thrust major
\begin{equation}
  \label{eq:TM}
T_M = \max_{\vec{n}}\frac{1}{Q}{\sum_h}'|\vec{p}_h\cdot\vec{n}|\>,
\qquad \vec{n}\cdot \vec{P}=0\>.
\end{equation}
The jet events we want to select with $p_t\sim Q$ have $T_M=\cO{1}$.
To select these events we prefer to use, instead of $T_M$, the
$(2+1)$-jet resolution variable $y_2$ defined by the $k_t$ jet
clustering algorithm \cite{DISkt} (see later). (By $(2+1)$-jet we mean
two hard outgoing jets in addition to the beam jet.)  To avoid small and large $T_M$
we require $y_{-} < y_2 <y_{+}$ with $y_{\pm}$ some fixed
values. Using this variable we have fewer hadronization corrections,
see \cite{DMW,Dpar}.

The observable $\Ko$ is similar to the out-of-plane jet shape studied
in $\ee$ annihilation \cite{Tmin} and in $Z_0$ production in hadron
collisions \cite{hh}. In the first case the event plane is defined by
the thrust and the thrust major axes. In the second case the event
plane is fixed by the beam axis and the $Z_0$ momentum.
The reason to analyse distributions in the out-of-plane momentum is
that the observable is sufficiently inclusive to allow an analytical
study.  The analysis of the in-plane momentum components is more
involved since one needs to start by fixing the jet rapidities.

Our analysis of $\Ko$ will make use of the methods introduced for the
study of the two observables in \cite{Tmin} and \cite{hh}.
In the present case we have one hadron in the initial state, so that
incoming radiation contributes both to the observable $\Ko$ and to the
parton density evolution.  To factorize these two contributions we
follow the method used in hadron-hadron collisions \cite{hh}.
The result is expressed in terms of the following factorized pieces:
\begin{itemize}
\item incoming parton densities obtained by resumming all terms
  $\as^n\ln^n\mu/\Ko$ ($\mu$ is the small factorization scale needed
  to subtract the collinear singularities and $\Ko$ is the hard scale
  for this distribution);
\item ``radiation factor'' characteristic of our observable. Its
  logarithm is obtained by resumming all DL and SL terms
  ($\as^n\ln^{n+1} \Ko/Q$ and $\as^n\ln^{n} \Ko/Q$ respectively). 
\item matching of the 
resummed result with the fixed order exact
  calculations \cite{DISENT,DISfix}. 
\end{itemize}
The factorization of the incoming parton densities and the radiation
factor is the crucial step for the present analysis. This result is
due to coherence and real/virtual cancellations (see \cite{hh} and
later). As a result, after this factorization procedure, the radiation
factor is a CIS quantity similar to the ones entering in the $\ee$
observables.

In order to make quantitative predictions we need to add to the above
PT result the $1/Q$--power corrections. We follow the procedure
successfully used in the analysis of $\ee$ jet-shape distributions
\cite{DMW}.  
The definition of the event plane makes our observable sensitive to
hard parton recoil. Here only the two final state hard partons can
take a recoil, while the initial state one is fixed along the beam
axis.  
This simplifies the treatment both of the PT
distribution and of the interplay between PT and NP effects, which is
a characteristic feature of all rapidity independent observables.

The $1/Q$ coefficient is expressed in terms of a single parameter,
$\al_0(\mu_I)$, given by the integral of the QCD coupling over the
region of small momenta $k_t\le \mu_I$ (the infrared scale $\mu_I$ is
conventionally chosen to be $\mu_I=2\GeV$, but the results are
independent of its specific value).  Effects of the non-inclusiveness
of $\Ko$ are included by taking into account the Milan factor $\cM$
introduced in \cite{Milan} and analytically computed in \cite{Milan2}.
The NP parameter $\al_0(\mu_I)$ is expected to be the same for all jet
shape observables linear in the transverse momentum of the emitted
hadrons. It has been measured only for $2$-jet event shapes and
appears to be universal with a reasonable accuracy
\cite{Exp-shape}. It is interesting to investigate if this
universality pattern also holds for near-to-planar $3$-jet
observables.

In order to improve the readability of the paper, in the main text we
present only the results and discuss their physical meaning, leaving
the detailed derivation of the results to a few technical appendices.
In section \ref{sec:Observable} we define the distribution and specify
the phase space region of $\Ko$ in which we perform the QCD study.
In section \ref{sec:QCD} we describe the PT and NP result for the
$\Ko$ distribution. We stress how the answer has a transparent
interpretation based on simple QCD (and kinematical) considerations.
In section \ref{sec:Matching} we improve our theoretical prediction by
performing the first-order matching and present some numerical
results.
Finally, section \ref{sec:Discussion} contains a summary, discussion
and conclusions.  

\section{The process and the observable \label{sec:Observable}}
We work in the Breit frame
\begin{equation}
  \label{eq:qP}
q = \frac{Q}{2}(0,0,0,2)\>, \qquad
P = \frac{Q}{2\xB}(1,0,0,-1) \>,\qquad
\xB = \frac{Q^2}{2(Pq)} \>,
\end{equation}
in which $2\xB P+q$ is at rest. Here $P$ and $q$ are the momenta of the
incoming proton and the exchanged vector boson ($\gam$ or $Z_0$).

In this frame, the rapidity (with respect to the direction of the
incoming proton) of an emitted hadron with momentum $p_h$ is given by
\begin{equation}
\label{eq:rapB}
\eta_h = \frac{1}{2}\ln\left(1+\frac{(qp_h)}{\xB (Pp_h)}\right).
\end{equation}
To avoid measurements in the beam region, the outgoing hadrons $p_h$
are taken in the rapidity range
\begin{equation}
\label{eq:rapidity}
\eta_h<\eta_0 \simeq
-\ln\tan\frac{\Theta_0}{2}\>,
\end{equation}
which corresponds to a cut of angle $\Theta_0$ around the beam
direction.  Similarly, the sums $\sum_h'$ in \eqref{eq:Kout} and
\eqref{eq:TM} extend over all hadrons with rapidities in the range
\eqref{eq:rapidity}.

To select jet events with $p_t\sim Q$ we use the $(2+1)$-jet resolution
variable $y_2$ introduced in the $k_t$-algorithm for DIS processes
\cite{DISkt}. For completeness we recall its definition.

Given the set of all outgoing momenta one defines the ``distance'' of
$p_h$ from the incoming proton momentum
\begin{equation}
y_{hP}=\frac{2}{Q^2}\,E^2_h(1-\cos\theta_{hP})\>.
\end{equation}
For any pair $p_{h'}$ and $p_{h''}$ of outgoing momenta one also defines
the ``distance''
\begin{equation}
y_{h'h''}=\frac{2}{Q^2}\,\min\{E^2_{h'},E^2_{h''}\}(1-\cos\theta_{h'h''})\>.
\end{equation}
If $y_{hP}$ is smaller than all $y_{h'h''}$, the hadron $p_h$ is
considered part of the beam jet and removed from the outgoing momentum
set.  Otherwise, the pair of momenta $p_{h'},p_{h''}$ with the minimum
distance $y_{h'h''}$ are substituted with the pseudo-particle (jet)
momentum $p_{h'''}=p_{h'}+p_{h''}$ ($E$-scheme).  The procedure is
repeated with the new momentum set until only two outgoing momenta
$p_{h_1},p_{h_2}$ are left. Then the final value of $y_2$ is defined
as
\begin{equation} 
\label{eq:y3} 
y_2\>\equiv\>\min\{y_{h_1P},y_{h_2P},y_{h_1h_2}\}\>.
\end{equation}
To select jet events with $p_t\sim Q$, as stated before, we require
\begin{equation}
  \label{eq:yc}
y_{-}<y_2<y_{+}\>,
\end{equation}
with $y_{\pm}$ fixed limits.  

The distribution we study is then defined as
\begin{equation}
  \label{eq:dsigma}
\begin{split}
  \frac{d\sigma(y_{\pm},\Ko)}{d\xB\,dQ^2}= \sum_{m}\int \frac{d\sigma_m}
  {d\xB\,dQ^2}\,\Theta(y_{+}\!-\!y_2) \Theta(y_2\!-\!y_{-}) 
\,\Theta\left(\Ko\!-\!\sum_{h=1}^m{\!}'\,|p_{h}^{\out}|\right),
\end{split}
\end{equation}
with $d\sigma_m/d\xB dQ^2$ the distribution for $m$ emitted hadrons in
the process under consideration.  Considering the cross section for
the $y_2$-range \eqref{eq:yc}
\begin{equation}
  \label{eq:sigma-yc}
\begin{split}
  \frac{d\sigma(y_{\pm})}{d\xB\,dQ^2}= \sum_{m}\int \frac{d\sigma_m}
  {d\xB\,dQ^2}\>\Theta(y_{+}\!-\!y_2) \Theta(y_2\!-\!y_{-})\>,
\end{split}
\end{equation}
we have the normalized distribution
\begin{equation}
  \label{eq:Sigma}
  \Sigma(y_{\pm},\Ko)=\frac{d\sigma(y_{\pm},\Ko)}{d\xB\,dQ^2}
\left\{\frac{d\sigma(y_{\pm})}{d\xB\,dQ^2}\right\}^{-1}.
\end{equation}
Due to the rapidity limitation \eqref{eq:rapidity} in the definition
of $y_2$, $\Ko$ and the event plane, the distribution will depend on
$\eta_0$. To avoid a strong $\eta_0$-dependence we will consider
$\eta_0$ and $\Ko$ in the range (see later)
\begin{equation}
  \label{eq:Komin}
  \Ko\>\ga \>
Q\,e^{-\eta_0}\>.
\end{equation}
The fact that this distribution is rich in information on the hard
process is clear from the fact that it depends not only on the
observable $\Ko$ but also on the variables $y_{\pm}$ which define
the geometry of the jet events with $p_t\sim Q$. In the range
\eqref{eq:Komin} the PT result does not depend on $\eta_0$, to our
accuracy. However, as we shall discuss, the power correction depends
linearly on $\eta_0$. This is due to the fact that the contributions
to the observable are uniform in rapidity.

\section{QCD result \label{sec:QCD}}
The QCD calculation of the distribution \eqref{eq:Sigma} is based on
the factorization \cite{DDT-BCM} of parton processes into the following
subprocesses:
\begin{itemize}
\item elementary hard distribution;
\item incoming parton distribution;
\item radiation factor corresponding to the observable $\Ko$.
\end{itemize}
These factors are described in the following subsections.

\subsection{Elementary hard process}
For the elementary hard vertex
\begin{equation}
  \label{eq:El-proc}
  q\,P_1\to P_2\,P_3\>,
\end{equation}
we introduce the kinematical variables (see Appendix \ref{App:El})
\begin{equation}
  \label{eq:El-kin} 
\isp=\frac{(P_1 P_2)}{(P_1 q)} \>,\qquad 
\ics = \frac{Q^2}{2(P_1 q)}>\xB \>.
\end{equation}
The invariant masses are
\begin{equation}
  \label{eq:Q_ab}
Q^2_{ab}=2(P_aP_b)\>,\quad
Q^2_{12}=\frac{\isp}{\ics}Q^2\>,\quad   
Q^2_{13}=\frac{1-\isp}{\ics}Q^2\>,\quad
Q^2_{23}=\frac{1-\ics}{\ics}Q^2\>.
\end{equation}
The thrust major for the process \eqref{eq:El-proc} is given by
$T_M=2\sqrt{\isp(1\!-\!\isp)(1\!-\!\ics)/\ics}$.
The substitution $\isp\to (1\!-\!\isp)$ interchanges $P_2$ and $P_3$,
so we distinguish $P_2$ from $P_3$ by assuming
\begin{equation}
  \label{eq:P2P3}
  (P_1P_2)\> < \>(P_1P_3)\>,
\end{equation}
which restricts us to the region $0<\isp<\half$. The variable $y_2$
for the elementary vertex is given in terms of the variables in
\eqref{eq:El-kin} by 
\begin{equation}
  \label{eq:El-y3}
\begin{split}
  y_2=\left\{
\begin{split}
&\>\>\isp^2+\isp(1-\isp)\frac{1-\ics}{\ics}
\>,\qquad\qquad\>\>\>\>\>\>\mbox{for}\>\>\ics<\frac{1+\isp}{1+2\isp}\>,
\\&\left(\frac{1-\ics}{\ics}\right)
\frac{\ics\isp+(1-\isp)(1-\ics)}{\ics(1-\isp)+\isp(1-\ics)}
\>,\quad\mbox{for}\>\>\ics>\frac{1+\isp}{1+2\isp}\>.
\end{split}
\right.
\end{split}
\end{equation}
Inverting this one finds $\isp=\isp(x,y_2)$. This function and the relative 
phase space are discussed in Appendix \ref{App:El}. 

We consider now the nature of the involved hard primary partons. We
identify the incoming parton of momentum $P_1$ by the index
$\can=q,\bar q,g$. Since \eqref{eq:P2P3} distinguishes $P_2$ from
$P_3$, in order to completely fix the configurations of the three
primary partons, we need to give an additional index $\conf=1,2,3$
identifying the gluon.  Therefore the primary partons with momenta
$\{P_1,P_2,P_3\}$ are in the following five configurations
\begin{equation}
\label{eq:configs}
\begin{split}
&\can=g,\,\conf=1\>\>\to\>\>\{gq\bar q\}\,,
\quad \mbox{or}\quad\{g\bar q q\}\,\\
&\can=q,\,\conf=2\>\>\to\>\>\{qgq\}\,, \\
&\can=\bar q,\,\conf=2\>\>\to\>\>\{\bar q g\bar q\}\,, \\
&\can=q,\,\conf=3\>\>\to\>\>\{qqg\}\,, \\
&\can=\bar q,\,\conf=3\>\>\to\>\>\{\bar q\bar q g\}\,.
\end{split}
\end{equation}
In Appendix \ref{App:El} we give the corresponding five elementary
distributions $d\hat\sigma_{\can,\conf,f}$, with $f$ the fermion
flavours.  The presence of the index $\can$ as well as $\conf$ is due
to the parity-violating term in the cross-section associated with
$Z_0$ exchange, such that the elementary cross-sections differ for
incoming quark and antiquark of the same flavour. If we consider only
photon exchange then the index $\can$ is redundant.

\subsection{Factorized QCD distribution}
The process \eqref{eq:ep} is described in QCD by one
incoming parton of momentum $p_1$ (inside the proton), and two
outgoing hard partons $p_2,p_3$ accompanied by an ensemble of
secondary partons $k_i$
\begin{equation}
  \label{eq:partonproc}
  q\,p_1 \to p_2\,p_3\,k_1\cdots k_n\>.
\end{equation}
Taking a small subtraction scale $\mu$ (smaller than any other scale
in the problem), we assume that $p_1$ (and the spectators) are
parallel to the incoming hadron,
\begin{equation}
  \label{eq:p1}
p_1=x_1 P\,.
\end{equation}
Therefore, the observable we study is
\begin{equation}
  \label{eq:Ko}
  \Ko=|p_{2x}|+|p_{3x}|+{\sum_i}' |k_{ix}|\>,
\end{equation}
where the $x$-axis corresponds to the out-of-plane direction, the
$z$-axis to the Breit direction, and the event plane is the
$y$-$z$ plane.  For large $y_2$ 
the hard partons $p_2,p_3$ are emitted at large angle. 
For small $\Ko$, $p_2,p_3$ and the secondary parton momenta $k_i$ are
near the event plane.

The event plane definition \eqref{eq:TM} corresponds to the
condition
\begin{equation}
\label{eq:ev-plane0}
p_{2x} + {\sum_U}'k_{ix} = p_{3x} + {\sum_D}'k_{ix}\>,
\end{equation}
which, together with momentum conservation, leads directly to
\begin{equation}
\label{eq:ev-plane}
p_{2x} = -{\sum_U}'k_{ix} - \half{\sum}''k_{ix}\>,\qquad
p_{3x} = -{\sum_D}'k_{ix} - \half{\sum}''k_{ix}\>.
\end{equation}
Here, by $U$ and $D$ we indicate the up- and down-regions corresponding 
to partons $k_i$ with $k_{iy}>0$ and $k_{iy}<0$ respectively.  Again, by
${\sum}'$ we indicate the sum restricted to secondary partons in the
region \eqref{eq:rapidity}.  By ${\sum}''$ we indicate the sum
restricted to secondary partons $k_i$ in the ``beam-region'' with
$\eta_i>\eta_0$.

We perform the QCD analysis at the accuracy required to make a
quantitative prediction: DL and SL resummation, matching with exact
fixed order results, and leading $1/Q$ power correction. The analysis
is similar to the one performed in \cite{Tmin,Dpar,hh}: the starting
point is the elementary process \eqref{eq:El-proc}. Then one considers
the secondary radiation (soft and/or collinear) in a $3$-jet
environment.  Finally one takes into account the exact matrix element
corrections and power corrections.

The application to the present case is described in detail in Appendix
\ref{App:Radiation} where we show that the distribution can be
expressed in the following factorized structure
\begin{equation}
\label{eq:QCD-dsigma}
\frac{d\sigma(\Ko,y_{\pm})}{d\xB\,dQ^2} = 
\sum_{\can,\conf,f}\int_{\xB}^{x_M}\frac{d\ics}{\ics}
\int_{\isp_{-}}^{\isp_+}d\isp
\left(\frac{d\hat\sigma_{\can,\conf,f}}{d\ics\,d\isp\,dQ^2}\right)
\cdot \cI_{\can,\conf,f}(\Ko,\ics,\isp,Q,\xB,\eta_0)\>,
\end{equation}
where $x_M$ is a function of $y_-$, the limits
$\isp_{\pm}=\isp_{\pm}(x,y_{\pm})$ select jet events with $p_t\sim Q$
and $d\hat\sigma$ are the distributions for the elementary hard
process \eqref{eq:El-proc}. See Appendix \ref{App:El} for the
elementary distributions and kinematics.

The distributions $\cI$, which resum higher order QCD emission, can be
expressed in the following factorized form
\begin{equation}
  \label{eq:Factor}
\cI_{\can,\conf,f}
\>=\>C_{\can,\conf}(\as)\cdot
\cP_{\can,f}\left(\frac{\xB}{\ics},\Ko\right)\cdot
\cA_{\conf}\left(\Ko,\ics,\isp,\,Q,\eta_0\right)\>.
\end{equation}
Here we describe the various factors:
\begin{itemize}
\item the factor $\cP_{\can,f}$ is the incoming parton distribution.
  It is given, for the various cases, by the quark, antiquark or gluon
  distribution inside the proton
\begin{equation}
  \label{eq:cP}
\begin{split}
&\cP_{q,f}(x,\Ko)\!=\!q_f(x,\Ko)\,,\quad
\cP_{\bar q,f}(x,\Ko)\!=\!{\bar q}_f(x,\Ko)\,,\\
&\cP_{g,f}(x,\Ko)\!=\!g(x,\Ko)\,.
\end{split}
\end{equation}
We show that here the hard scale is fixed at $\Ko$ and that the
dependence on $\eta_0$ can be neglected as long as we take $\Ko$
sufficiently large in the range \eqref{eq:Komin};
\item the distribution $\cA_{\conf}$ is the CIS radiation factor
  which resums powers of $\ln\Ko/Q$
  and is a CIS quantity. It is sensitive only to QCD radiation and
  therefore does not depend on the flavour (we neglect quark masses).
  There are various hard scales in $\cA_{\conf}$ (given in terms of the
  $Q_{ab}^2$ in \eqref{eq:Q_ab}) which are determined by the
  SL accuracy analysis. This quantity is similar to $3$-jet shape
  distributions one encounters in $\ee$ annihilation processes
  \cite{Tmin,Dpar} and in hadron collisions \cite{hh};
\item the first factor is the non-logarithmic coefficient function
  with the expansion
\begin{equation}
  \label{eq:Coeff}
C_{\can,\conf}(\as)=1+c_1\,\frac{\as}{2\pi}+ \cO{\as^2}\,, 
\quad \as=\as(Q)\>.
\end{equation}
It takes into account hard corrections not included in the other two
factors and is obtained from the exact fixed order results.
\end{itemize}

The factorization of the first two pieces is based on the fact that
contributions to $\cA$ (to $\cP$) come from radiation at angles larger
(smaller) than $\Ko/Q$.  This implies that one is able to reconstruct
the parton densities $\cP$ as for the DIS total cross sections in
which one does not analyse the emitted radiation. The only difference
here is that the parton density hard scale is given by $\Ko$ while in
the fully inclusive case of DIS total cross section the hard scale is
$Q$.  As a result of this factorization, the radiation factor $\cA$ can
be analysed by the same methods used in $\ee$.

The basis for the factorized result \eqref{eq:Factor} has been
discussed in \cite{hh} and it is based on coherence and real/virtual
cancellations.  Since this is a crucial point for our analysis we
recall in some detail the relevant steps.

\begin{itemize}
\item Since $\Ko$ is a CIS (global~\cite{nonglobal}) quantity 
(its value does not change if one of the emitted particles branches
into collinear particles or undergoes soft brems\-strah\-lung), within
SL accuracy we can systematically integrate over the final state
collinear branchings.
\item Factorizing the phase space and the observable (see
  \eqref{eq:dPhi}) by Mellin and Fourier transforms one has that each
  secondary parton of momentum $k_i$ contributes with an inclusive
  factor
\begin{equation}
    \label{eq:I}
    I(k_i)=1-\eps(k_i)\cdot U(k_i)\,,
\end{equation}
where $1$ is the virtual term while the factors $\eps$ and $U$ are the
contribution from the real emission. The factor $\eps(k_i)$ accounts
for the energy loss of the primary incoming parton $p_1$ due to
emission of collinear secondary partons.  Therefore, for $k_i$
non-collinear to $p_1$ we have $\eps(k_i)=1$, while for $k_i$ collinear
to $p_1$ we have $\eps(k_i)=z_i^{N-1}$ with $1-z_i$ the 
energy fraction of $k_i$ with respect to $p_1$ and $N$ the usual Mellin
moment of the anomalous dimension, see \eqref{eq:eps}.  The factor
$U(k_i)$ depends on the observable and in our case is given by
\eqref{eq:U}.

\item The crucial point which is the basis of the factorization in
  \eqref{eq:Factor} is that, to SL accuracy for $\Ko\ll Q$, we can
  replace (see \eqref{eq:rho})
\begin{equation}
  \label{eq:theta}
  [1-U(k_i)]\simeq \Theta(k_{ti}-k_0)\>,\qquad k_0\sim \Ko\>,
\end{equation}
so that we can write
\begin{equation}
    \label{eq:I'}
I(k_i) \simeq \Theta\left(k_{ti}\!-\!k_0\right)+
\left[1\!-\!z_i^{N-1}\right]\Theta\left(k_0-k_{ti}\right)\>.
\end{equation}
The first term contributes to the radiation factor while the second
reconstructs the anomalous dimensions for the various channels. The two
contributions do not interfere and, as discussed above, one obtains
the factorized expression \eqref{eq:Factor}.
\end{itemize}

In the next two subsections we will describe our result for the
radiation factor $\cA_{\conf}$. First we describe the PT contribution
obtained by resumming the logarithmically enhanced terms at SL
accuracy. Then we describe the leading NP corrections originating from
the fact that the (virtual) momentum in the argument of the running
coupling cannot be prevented from vanishing, even in hard
distributions. Matching with the exact first order result is
considered later.

\subsection{The PT radiation factor and distribution}
The PT radiation factor $\cA_{\conf}^{\PT}$ can be expressed as (see
Appendix \ref{App:Radiation})
\begin{equation}
\label{eq:PT-cA}
\cA_{\conf}^{\PT}
\>=\> e^{-R_{\conf}\left(\Ko^{-1},\ics,\isp,Q\right)}
\cdot \cS_{\conf}\left(\as\ln\frac{Q}{\Ko}\right)\>.
\end{equation}
It resums DL and SL contributions originating from the emission of
soft or collinear secondary partons $k_i$ in \eqref{eq:partonproc}.
The PT radiator $R_{\conf}$ in the first factor contains the DL
resummation together with SL contributions coming from the running
coupling and the proper hard scales.  The second factor is the SL
function which resums effects due to multiple radiation, including hard
parton recoil.  For $\Ko$ in the region \eqref{eq:Komin}, this PT
result does not depend on $\eta_0$.  The analysis leading to this
result is very similar to that for other $3$-jet distributions
\cite{Tmin,Dpar,hh} (see Appendix \ref{App:RadPT}).  Here we report
the relevant expressions and illustrate their physical aspects.

The PT radiator $R_{\conf}$ is given by a sum of three contributions
associated with the emission from the three primary partons $P_a$. 
For the configuration $\conf$, it is given by
\begin{equation}
  \label{eq:PT-rad}
R_{\conf}\>=\>\sum_{a=1}^3 C^{(\conf)}_a\,
r\left(\Ko^{-1},\zeta_a^{(\conf)} Q_a^{(\conf)}\right)\>,
\end{equation}
where
\begin{equation}
  \label{eq:QCs}
\begin{split}
&Q_{a}^{(\conf)}=Q_{b}^{(\conf)}=Q_{ab}\>,%a,b\neq\conf
\qquad
Q_{\conf}^{(\conf)}=\frac{Q_{a\conf}Q_{\conf b}}{Q_{ab}}\>,\\
&C_{a}^{(\conf)}=C_{b}^{(\conf)}=C_F\>,\qquad\>\>\>
C_{\conf}^{(\conf)}=C_A\>,\\
&\zeta_{a}^{(\conf)}=\zeta_{b}^{(\conf)}=e^{-\frac{3}{4}}\>,\qquad\>\>\>\>
\zeta_{\conf}^{(\conf)}=e^{-\frac{\beta_0}{4N_c}}\>,
\end{split}
\end{equation}
with indices $a,b$ corresponding to the fermions and $\conf$ to
the gluon, i.e.~$a\ne b\ne\conf$.

As is typical for a $3$-jet quantity, the scale for the quark or
antiquark terms of the radiator is the fermion invariant mass,
while the scale for the gluon term is given by the gluon transverse
momentum with respect to the fermion system. The rescaling
factors $\zeta_{a}^{(\conf)}$ take into account SL contributions from
non-soft secondary partons collinear to the primary partons $P_a$.

In \eqref{eq:PT-rad}
$r$ is the following DL function
\begin{equation}
\label{eq:r}
r\left(\Ko^{-1},Q'\right)\equiv 
\int_{\Ko}^{Q'}\frac{dk}{k}\frac{2\as(2k)}{\pi}
\ln\frac{Q'}{2k}\>,
\end{equation}
and $k$ is the out-of-plane component of momentum and $\as$ is taken
in the physical scheme \cite{CMW}. The rescaling factor $2$ in the
running coupling comes from the integration over the in-plane momentum
component. The fact that $\Ko$ is the lower bound in the PT radiator,
to this order, comes from real/virtual cancellation (see \eqref{eq:theta}
and \eqref{eq:rho}).  The exact expression of the hard scales and the
rescaling factors in \eqref{eq:QCs} is relevant at SL level.

The factor $\cS$ is expressed in terms of the SL function
\begin{equation}
  \label{eq:r'}
r'=r'(\Ko^{-1},Q) \equiv \frac{2\as(\Ko)}{\pi}\>\ln\frac{Q}{\Ko}\>,
\end{equation}
given by the logarithmic derivative of $r$, apart from terms beyond SL
accuracy. We find
\begin{equation}
  \label{eq:SL}
\cS_{\conf}=\frac{e^{-\gam_E\,C_T\,r'}}{\Gamma(1+C_T\,r')}\cdot 
\cF\left(C_{12}^{(\conf)}\,r'\right)
\cF\left(C_{13}^{(\conf)}\,r'\right),
\qquad C_T=2C_F+C_A\>,
\end{equation}
where $C_T$ is the total colour charge of the primary partons (see
\eqref{eq:El-proc}). The first factor is the same for all $3$-jet shape
emission distributions. The second factor depends on the specific
observable. The function $\cF$ is given by
\begin{equation}
  \label{eq:cF}
\cF(z)=\frac{\Gamma(\frac{1+z}{2})}
{\sqrt{\pi}\Gamma(1+\frac{z}{2})}\>.
\end{equation}
It takes into account the correct kinematics for the emission and in
particular the effect of the recoil of the two outgoing primary
partons in \eqref{eq:El-proc} due to the emission of secondary QCD
radiation.  Here $C_{1a}^{(\conf)}$ is given by the colour charge of
the outgoing parton $a=2,3$ plus half of the charge of the incoming
parton, in the given configuration,
\begin{equation}
 \label{eq:C1a}
C_{1a}^{(\conf)}=\half C_1^{(\conf)}+C_a^{(\conf)}\>.
\end{equation}
This combination of colour charges enters due to the kinematics that
defines the event plane (see \eqref{eq:ev-plane}), which leads to the
vanishing of the sum of the out-of-plane momenta in the ``up-region'',
i.e. with positive $y$-components, and also in the ``down-region''.
(We are working in the r\'egime in which the effect of the rapidity
cut is negligible.)  The momenta of outgoing primary partons $\#2$ and
$\#3$ are in the up- and down-region respectively. The incoming parton
$\#1$ is instead along the Breit axis and emits equally into both
regions.  Notice that $C_{12}^{(\conf)}+C_{13}^{(\conf)}=C_T$.

Finally the DL and SL resummed PT part of the distribution $\cI$ is
obtained from \eqref{eq:Factor} by using the radiation factor in
\eqref{eq:PT-cA}
\begin{equation}
  \label{eq:PT-cI}
\cI^{\PT}_{\can,\conf,f}
\>=\>C_{\can,\conf}(\as)\cdot
\cP_{\can,f}\left(\frac{\xB}{\ics},\Ko\right)\cdot
\cA^{\PT}_{\conf}\left(\Ko,\ics,\isp,Q\right)\>.
\end{equation}
The coefficient function will be computed at one-loop by using the
numerical program DISENT of Ref.~\cite{DISENT} and subtraction of the
one-loop contribution already contained in the two factors $\cP$ and
$\cA$.

To first order in the PT expansion we have
\begin{equation}
  \label{eq:oneloop} 
\frac{d\cA^{\PT}_{\conf}}{d\ln\Ko}=\frac{2\as(Q)}{\pi}
\sum_{a=1}^3 C_a^{(\conf)}
\ln\frac{\zeta_a^{(\conf)} Q_a^{(\conf)}}{\Ko}+\cO{\as^2}\>.
\end{equation}
The absence of the factor $\half$ in the logarithm compared with that
of equation \eqref{eq:r} is due to the event-plane kinematics (see
\eqref{eq:ev-plane}). For a single gluon $k_x$ (emitted from $p_1,p_2$
or $p_3$) only one of the two hard partons $p_2$ or $p_3$ takes recoil
equal to $|p_{ax}|=|k_x|=\Ko/2$.

\subsection{The distribution including NP corrections \label{sec:NPshift}}
As in other cases of jet-shape distributions, the leading NP
correction corresponds to a {\it shift} in the PT distribution. This
is due to the fact that, in the Mellin representation of $\cI$, the PT
radiator is affected by a NP correction with leading term {\it linear}
in the Mellin variable. Thus its effect corresponds to a shift in the
conjugate variable, $\Ko$, and one has
\begin{equation}
  \label{eq:shift}
\cI_{\can,\conf,f}(\Ko)=
\cI^{\PT}_{\can,\conf,f}\left(\Ko-\delta \Ko^{(\conf)}\right).
\end{equation}
The quantity $\delta \Ko^{(\conf)}$, which can be cast in the
following form
\begin{equation}
  \label{eq:shift-tot}
 \begin{split}
\de \Ko^{(\conf)}=\cp\,\left\{\frac{C_{12}^{(\conf)}}{C_T}\cdot
\sum_{b=1}^3C_b^{(\conf)}\,Y^{(\conf)}_{2b}
\>+\>\frac{C_{13}^{(\conf)}}{C_T}\cdot
\sum_{b=1}^3C_b^{(\conf)}\,Y^{(\conf)}_{3b}\right\}\>,
 \end{split} 
\end{equation}
corresponds to the integral over the infrared region of the soft gluon
distribution with weight $|k_x|$, the contribution to $\Ko$ from the
very soft gluon generating the NP contribution.  The first factor
$\cp$ corresponds to the momentum integral (including $|k_x|$ and the
running coupling), while $Y^{(\conf)}_{ab}$ is the rapidity interval
(more precisely the logarithmic integral of the angle which the very
soft gluon forms with the event plane).

The parameter $\cp$, given in \eqref{eq:cp}, is expressed in terms of
the NP parameter $\al_0(\mu_I)$ and the Milan factor $\cM$
\cite{Milan,Milan2} which takes into account effects of the
non-inclusiveness of $\Ko$. $\cp$ also contains renormalon
cancellation terms. The quantity $\al_0(\mu_I)$ is the integral of the
running coupling in the infrared region
\begin{equation}
  \label{eq:al0}
  \al_0(\mu_I)=\int_0^{\mu_I}\frac{dk_t}{\mu_I}\as(k_t)\>.
\end{equation}
The infrared scale $\mu_I$ is conventionally chosen to be
$\mu_I=2\GeV$, but the results are independent of its specific value.
Both $\al_0(\mu_I)$ and $\cM$ are the same for
all jet shape observables linear in the transverse momentum of emitted
hadrons. 

The derivation of the rapidity intervals $Y^{(\conf)}_{ab}$ is
reported in Appendix \ref{App:Radiation}, see \eqref{eq:Ys}. 
The expressions for $b=1$ are simply given by 
\begin{equation}
\label{eq:Ya1}
Y_{21}^{(\conf)}=Y_{31}^{(\conf)}=
\ln\frac{\ics e^{\eta_0}Q_1^{(\conf)}}{Q}\,.
\end{equation}
For the diagonal cases $a=b=2,3$ we have the behaviour 
\begin{equation}
  \label{eq:Ydiag}
Y_{22}^{(\conf)}\simeq\ln\frac{2Q_2^{(\conf)}\zeta}{\Ko}+\cO{r'}\>,\qquad
Y_{33}^{(\conf)}\simeq\ln\frac{2Q_3^{(\conf)}\zeta}{\Ko}+\cO{r'}\>,
\end{equation}
with $r'$ given in \eqref{eq:r'}, $Q^{(\conf)}_b$ the hard scales
in \eqref{eq:QCs} and $\zeta=2e^{-2}$ the rescaling factor which
accounts for the mismatch in the integration over an angle between two
vectors and a vector and a plane.  The expressions for the
off-diagonal cases have more complex behaviours. We have to
distinguish two regions.  For $\as\ln^2 Q/\Ko\ll 1$ we have (see
\eqref{eq:Laexp2a})
\begin{equation} 
\label{eq:many}
\begin{split}
Y_{23}^{(\conf)}\simeq
\frac{\pi}{2\sqrt{C_{13}^{(\conf)}\,\as(Q)}}\>,\qquad
Y_{32}^{(\conf)}\simeq\frac{\pi}{2\sqrt{C_{12}^{(\conf)}\,\as(Q)}}\>,
\end{split}
\end{equation}
with corrections of $\cO{1}$.
For $\as\ln^2 Q/\Ko\gg1$, we have (see \eqref{eq:Laexp1a})
\begin{equation}
\label{eq:few}
Y_{32}^{(\conf)}\simeq\ln\frac{2Q_2^{(\conf)}\zeta}{\Ko}+\cO{r'}\>,\qquad
Y_{23}^{(\conf)}\simeq\ln\frac{2Q_3^{(\conf)}\zeta}{\Ko}+\cO{r'}\>.
\end{equation}

Before illustrating and discussing these expressions, we observe the
following features of the distribution obtained in \eqref{eq:shift}.
Since $\delta \Ko^{(\conf)}$ depends on $\Ko$, the PT distribution is
actually deformed.
The dependence on $\eta_0$, the available rapidity interval for the
measurement, enters only through the NP shift (the PT distribution is
independent of $\eta_0$, to our accuracy, as long as we take $\Ko$ in
the region \eqref{eq:Komin}).
The dependence of $\delta \Ko^{(\conf)}$ on $y_2$ (through
$Q_{b}^{(\conf)}$), implies that the deformation of the PT
distribution differs for different geometries of the process.
To this order (leading $1/Q$ power) we can neglect NP corrections
arising from the anomalous dimension \cite{DMW} which are of
subleading order $1/Q^2$. 
Finally, we observe that the presence of $\ln\Ko$ and $1/\sqrt{\as}$
contributions to the shift are features common to the case of rapidity
independent observables such as, broadening \cite{broad}, $T_m$
\cite{Tmin} or the out-of-plane momentum in hadronic collisions
studied in \cite{hh}.

We now discuss how the above expressions for the rapidity intervals
$Y^{(\conf)}_{ab}$ are based on simple QCD considerations and on the
kinematics of the event-plane (see \eqref{eq:ev-plane}).
In the following we focus on the term \eqref{eq:shift-tot} 
\begin{equation}
  \label{eq:12b}
\frac{C^{(\conf)}_{12}}{C_T}\cdot C_b^{(\conf)}\cdot Y^{(\conf)}_{2b}\>,
\end{equation}
the other is obtain by exchanging $2\!\leftrightarrow\!3$.  It
corresponds to the contribution in which a PT gluon is emitted in the
up-region (this happens with probability $C^{(\conf)}_{12}/C_T$) and a
very soft gluon, generating the NP correction, is emitted off the
primary parton $b=1,2,3$ (with colour factor $C_b^{(\conf)}$).  For
this contribution we consider the various cases.

Case $b=1$. The incoming parton $p_1$, emitting the very soft
  gluon, is never affected by recoil since it is aligned along the
  Breit axis. Therefore the rapidity interval $Y_{21}$ is fixed by
  the boundary $\eta_0$, the available rapidity interval for the
  measurement (see \eqref{eq:rapidity}). The remaining term $\ln\ics
  Q_1^{(\conf)}/Q$ is related to the standard parton $\#1$ hard scale
  and a boost from the Breit frame of full process to the hard
  elementary vertex, see \eqref{eq:cut1a};

Case $b=2$.  The outgoing parton $p_2$ and the PT gluon are in
  the same region (the up-region) so that, from the event plane
  kinematics \eqref{eq:ev-plane}, $p_2$ undergoes recoil with
  $|p_{2x}|=\Ko/2$.  Since $\ln(\zeta Q^{(\conf)}_2/|p_{2x}|)$ fixes
  the rapidity interval for the very soft gluon emitted by $p_2$, we
  obtain the result \eqref{eq:Ydiag};

Case $b=3$. Here we cannot be limited to considering the contribution
  from a single PT gluon emitted in the up-region since, from the
  event plane kinematics \eqref{eq:ev-plane}, one would have
  $p_{3x}=0$ and the rapidity interval $Y_{23}$ would diverge (for a
  zero mass gluon). We then need to consider high-order PT
  contributions which push parton $\#3$ off the plane by a certain
  $p_{3x}$. The distribution in the recoil momentum $p_{3x}$ has two
  different r\'egimes:
  \begin{itemize}
  \item for $\as\ln^2 Q/\Ko\ll 1$ there are few secondary partons and
    the PT distribution is given by Sudakov form factors.  Integrating
    $\ln |p_{3x}|$, the rapidity interval contribution, we obtain the
    behaviour in \eqref{eq:many} (here $C^{(\conf)}_{13}$ is the
    coefficient of the DL exponent of the $|p_{3x}|$ Sudakov factor);
  \item for $\as\ln^2 Q/\Ko\gg 1$ the PT radiation is well developed,
    so that $p_3$ takes recoil with $|p_{3x}|\sim\Ko$ and we obtain the
    behaviour \eqref{eq:few}.
\end{itemize}

\section{Matching and numerical analysis \label{sec:Matching}}
The final step needed to obtain a quantitative prediction is the
calculation of the non-logarithmic coefficient function
\eqref{eq:Coeff} from the exact matrix element results.
We use the numerical program DISENT of Ref.~\cite{DISENT}, which
includes only the $\gamma$ exchanged contribution, and we compute the
first order coefficient $c_1$ by subtracting the DL and SL terms
already contained in the PT resummed result. The $Z_0$ contribution
will be important at large values of $Q$. For example, for $Q=20\GeV$
the $Z_0$ contribution to the cross section is of order $0.5\%$ (see
\eqref{eq:Z-cross}).

Taking into account only $\gamma$ exchange simplifies considerably the
expressions since we only need $\conf$ to identify the configuration,
while $\can$ becomes redundant. 
The distribution \eqref{eq:dsigma} and the cross-section for
the high $p_t$ jet \eqref{eq:sigma-yc} are given by
\begin{equation}
\label{eq:dsigma-gamma}
\begin{split}
\frac{d\sigma^{(\gam)}(y_{\pm},\Ko)}{d\xB\,dQ^2} = 
&\sum_{\conf,f}\int_{\xB}^{x_M}\frac{d\ics}{\ics}
\int_{\isp_{-}}^{\isp_{+}}d\isp
\left(\frac{d\hat\sigma_{\conf,f}^{(\gam)}}{d\ics\,d\isp\,dQ^2}\right)
\cdot \cI_{\conf,f}^{(\gam)}(\Ko,\ics,\isp,\,Q,\xB,\eta_0)\>,\\
\frac{d\sigma^{(\gam)}(y_{\pm})}{d\xB\,dQ^2} = 
&\sum_{\conf,f}\int_{\xB}^{\ics_M}\frac{d\ics}{\ics}
\int_{\isp_-}^{\isp_+}d\isp
\left(\frac{d\hat\sigma_{\conf,f}^{(\gam)}}{d\ics\,d\isp\,dQ^2}\right)
\cdot\cP_{\conf,f}^{(\gam)}\left(\frac{\xB}{x},Q\right)\>.
\end{split}
\end{equation}
where $x_M$ and $\isp_{\pm}=\isp_{\pm}(x,y_{\pm})$ are
found in Appendix \ref{App:El}. The hard elementary distribution is
given by (see Appendix \ref{App:El})
\begin{equation}
\label{eq:dsigma-gamma1}
\begin{split}
&\frac{d\hat{\sigma}_{\conf,f}^{(\gam)}}{d\ics\,d\isp\,dQ^2}= 
\frac{\alpha^2\as}{Q^4}e_f^2
\left[\left(2\!-\!2y\!+\!y^2\right)\,C_T^{\conf}(\ics,\isp)
+2(1\!-\!y)\,C_L^{\conf}(\ics,\isp)\right]\>,\\
&C_T^{3}=C_F\left[\frac{\ics^2\!+\!\isp^2}{(1\!-\!\ics)(1\!-\!\isp)}
+2(1\!+\!\ics\isp)\right]\>,\quad
C_T^{1} =\left[\ics^2+(1\!-\!\ics)^2\right]
\frac{\isp^2+(1\!-\!\isp)^2}{\isp(1-\isp)}\>,\\
&C_L^{3}= C_F\cdot 4\ics\isp \>, \qquad C_L^{1} = 8\ics(1\!-\!\ics)\>,
\qquad C_{T/L}^{2}(\ics,\isp)\!=\!C_{T/L}^{3}(\ics,1\!-\!\isp)\>,
\end{split}
\end{equation}
where $y=Q^2/(\xB s)$ with $\sqrt{s}$ the centre of mass energy of 
the process.

We start with the resummed PT result $\cI^{\PT}$ \eqref{eq:PT-cI}
which in this case takes the form
\begin{equation}
\label{eq:cI-gammaPT}
\cI_{\conf,f}^{(\gam)\>\PT}
\>=\>C_{\conf}^{(\gam)}(\as)\cdot
\cP_{\conf,f}^{(\gam)}\left(\frac{\xB}{\ics},\Ko\right)\cdot
\cA^{\PT}_{\conf}\left(\Ko,\ics,\isp,\,Q,\xB\right)\>,
\end{equation}
with the incoming parton distribution $\cP_\conf^{(\gam)}$ given by
\begin{equation}
\begin{split}
\cP_{\conf,f}^{(\gam)}(x,\Ko)=\left\{
\begin{split}
& g(x,\Ko)\>,\qquad\qquad\qquad\>\>\>\>\conf=1 \\
& q_f(x,\Ko)+\bar{q}_f(x,\Ko)\>,\>\>\>\conf=2,3\>.
\end{split}
\right.
\end{split}
\end{equation}
The PT radiation factor $\cA^{\PT}_\conf$ is the same whether we
include the $Z_0$ or not, and is given in \eqref{eq:PT-cA}.

\subsection{Matching}
We now discuss the first order matching.  To determine $c_1^{(\gam)}$,
the first order term of the coefficient function $C_{\conf}^{(\gam)}$
(see \eqref{eq:Coeff}), we expand the normalized resummed distribution
to one loop
\begin{equation}
\label{eq:Sigma-oneloop}
\begin{split}  
&\Sigma^{(\gam)}(y_{\pm},\Ko)=
\frac{d\sigma^{(\gam)}(y_{\pm},\Ko)}{d\xB\,dQ^2}
\left\{\frac{d\sigma^{(\gam)}(y_{\pm})}{d\xB\,dQ^2}\right\}^{-1}\\
&=1+\frac{\as(Q)}{2\pi}\!\left(G_{12}\ln^2\frac{Q}{\Ko}\!+\!
G^{(\gam)}_{11}(y_{\pm})\ln\frac{Q}{\Ko}+c^{(\gam)}_1(y_{\pm},\Ko)\right)
+\ldots 
\end{split}
\end{equation}
and compare with the result provided by the numerical program DISENT
\cite{DISENT}. From the resummed result we have $G_{12}=-2C_T$ and
\begin{equation}
\label{eq:G11}
\begin{split}
G^{(\gam)}_{11}(y_{\pm})\>&\frac{d\sigma^{(\gam)}(y_{\pm})}{d\xB\,dQ^2}
\>=\>\sum_{\conf,f}\int_{\xB}^{x_M}\frac{d\ics}{\ics}
\int_{\isp_{-}}^{\isp_{+}}d\isp
\left(\frac{d\hat\sigma_{\conf,f}^{(\gam)}}{d\ics\,d\isp\,dQ^2}\right)\\
&\cdot \left\{-4\sum_a C_a^{(\conf)}\,
\ln\frac{\zeta_a^{(\conf)}Q_a^{(\conf)}}{Q}\>
-\frac{2\pi}{\as(Q)}\frac{\partial}{\partial\ln Q}\right\}\cdot
\cP_{\conf,f}^{(\gam)}\left(\frac{\xB}{\ics},Q\right)\>.
\end{split}
\end{equation}
We have verified that the above coefficients $G_{12}$ and
$G^{(\gam)}_{11}$ are correctly reproduced by DISENT and by
subtraction we deduce the non-logarithmic one loop coefficient
$c^{(\gam)}_1(y_{\pm},\Ko)$.

We use the first-order (Log R)-matching prescription and write
\begin{equation}
  \label{eq:c1-gamma}
\left. C^{(\gam)}_{\conf}(\as)\right|_{\mat}=
e^{\frac{\as}{2\pi}c_1(y_{\pm},\Ko)}\>,
\end{equation}
with $\as=\al_{\MSbar}(Q)$.
We also implement the correct kinematical bound for $\Ko<\Ko^M$ (with
$\Ko^M$ obtained from DISENT) by replacing $\Ko$ by $\Ko'$, implicitly
defined by
\begin{equation}
\label{eq:Kotilde}
\frac{Q}{\Ko}\to\frac{Q}{\Ko'}
=\frac{Q}{\Ko}-\frac{Q}{\Ko^M}+1\>,
\end{equation}
in the final PT result \eqref{eq:cI-gammaPT} so that the differential
distribution vanishes for $\Ko=\Ko^M$.  Finally we make the
replacement
\eqref{eq:shift}, \eqref{eq:shift-tot} in order to include the NP
correction.

\subsection{Numerical results}

To summarize, the QCD prediction for the normalized distribution
$\Sigma(y_{\pm},\Ko)$ in \eqref{eq:Sigma} is obtained (for exchanged
$\gam$) from the ratio of the two distributions in
\eqref{eq:dsigma-gamma} in which the integrand $\cI$ is given by
\eqref{eq:cI-gammaPT} with the coefficient function given in
\eqref{eq:c1-gamma} and using the substitution
\eqref{eq:Kotilde} of $\Ko$. The radiation factor $\cA$ is given, at
PT level, by \eqref{eq:PT-cA}. Leading power corrections are included
by performing the substitutions in \eqref{eq:shift} and
\eqref{eq:shift-tot}.
 
We now report some plots for $d\Sigma(y_{\pm},\Ko)/d\Ko$ (data are not
yet available) for $s=98400\GeV^2$ and the rapidity cut $\eta_0=3$.  A
smaller rapidity cut, i.e. a larger value of $\eta_0$, would have two
advantages: the PT resummation is valid for smaller $\Ko$, and the NP
shift becomes larger; but this needs to be balanced against the
experimental resolution and the need to exclude the proton remnant.

The QCD predictions are all obtained for the following choices: the
parton density functions in \cite{SF}, set MRST2001; the NP parameter
$\al_0(\mu_I)=0.52$, a value in the range determined by the analysis
of $2$-jet observables in $\ee$ annihilation~\cite{GZ} and the running
coupling $\as(M_Z)=0.119$.

\EPSFIGURE[ht]{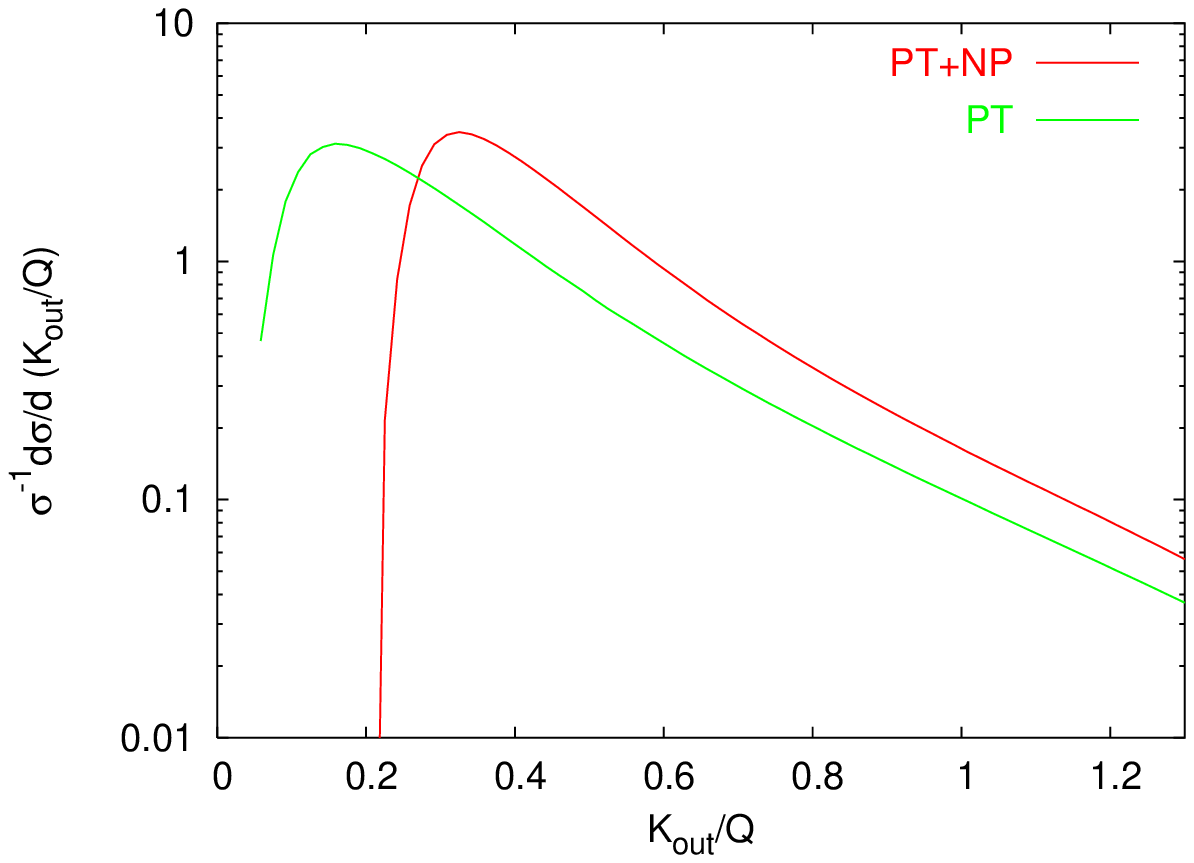,width=0.7\textwidth}{ 
QCD prediction for $\xB=0.1$, $Q^2=400\GeV^2$ and $y_-=0.1$.   
In the NP correction we have taken $\al_0(2\GeV)=0.52$. 
\label{fig:Q=20}}
 
\EPSFIGURE[ht]{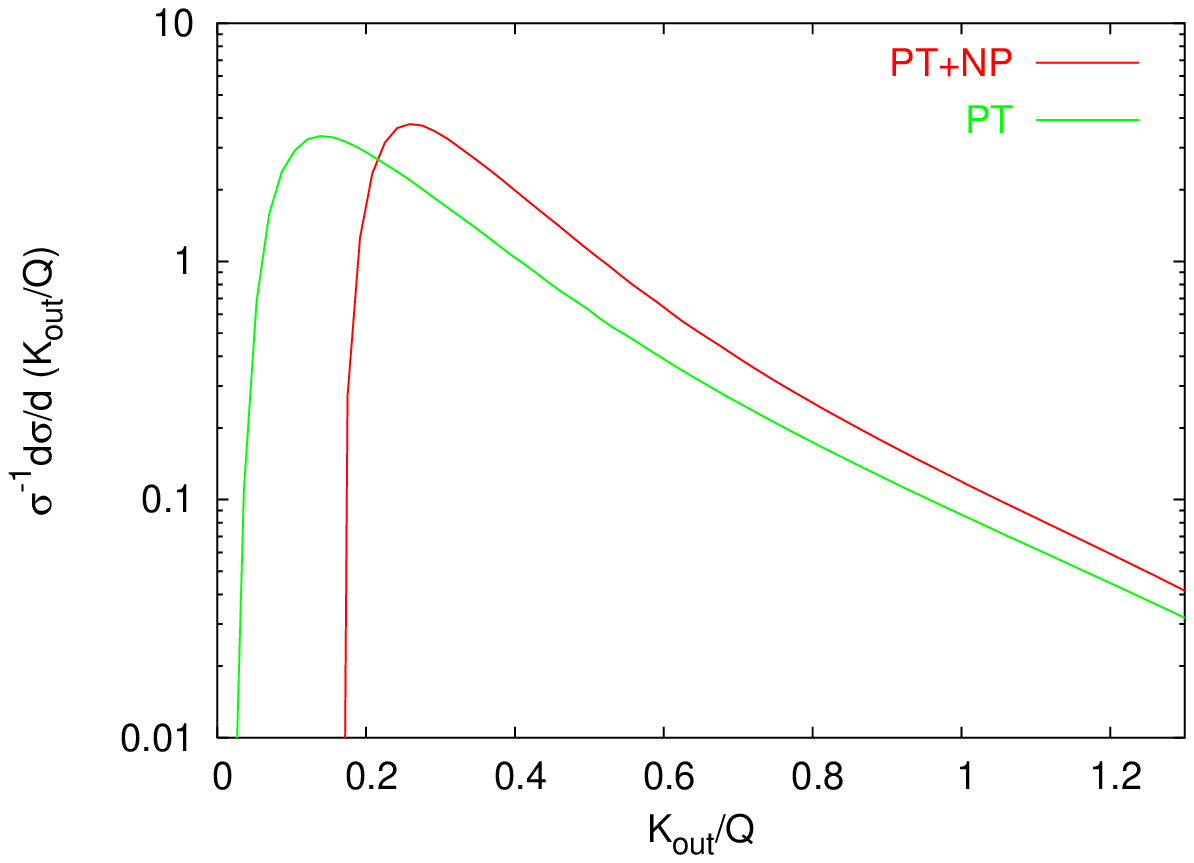,width=0.7\textwidth}
{The same as Fig.\ref{fig:Q=20} but with
$Q^2=900\GeV^2$.\label{fig:Q=30}}

Figs.~\ref{fig:Q=20} and \ref{fig:Q=30} show the differential $\Ko$
distribution for $Q=20\GeV$ and $Q=30\GeV$, and $\xB=0.1$. In order to
prevent $T_M$ from being too large it is enough to let $y_+$ go to its
maximum kinematically allowed value $y_+=2.5$ (see \eqref{eq:TMrange}
and \eqref{eq:isp_pm}). From these pictures one notices the $1/Q$
dependence of the power correction. One can also observe that the
shift is larger at smaller $\Ko$, so that the distribution is
squeezed. This is due to the $\ln \Ko$ dependence of the NP
correction (see discussion in section \ref{sec:NPshift}).

\EPSFIGURE[ht]{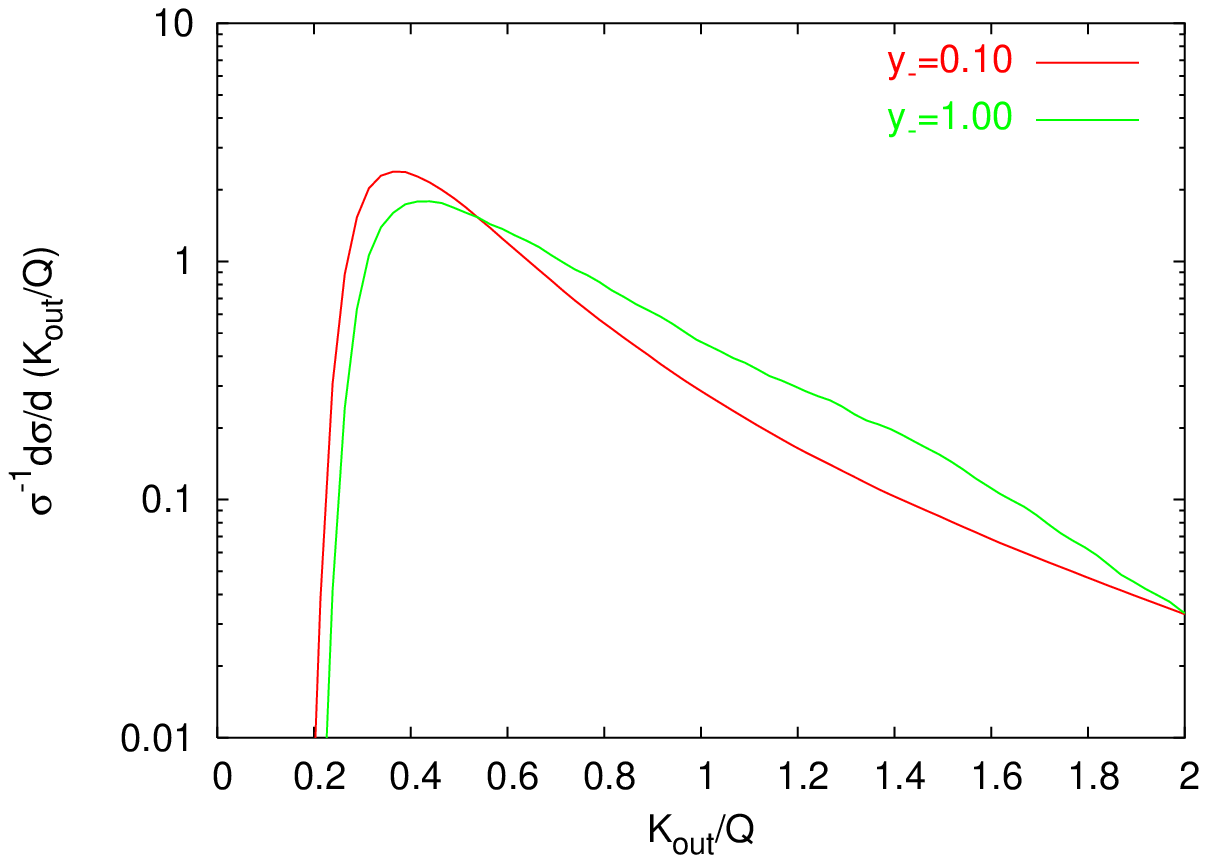,width=0.7\textwidth}
{The full (PT+NP) differential $\Ko$ distribution for two different
values of $y_-$, for fixed $y_+=2.5$, $\xB=0.01$ and
$Q^2=400\GeV^2$.\label{fig:yc}}

The fact that our predictions are also geometry dependent can be seen from
Fig.~\ref{fig:yc}, where we plot the PT+NP curve for two different
values of $y_-$, fixing $y_+=2.5$. 

\EPSFIGURE[ht]{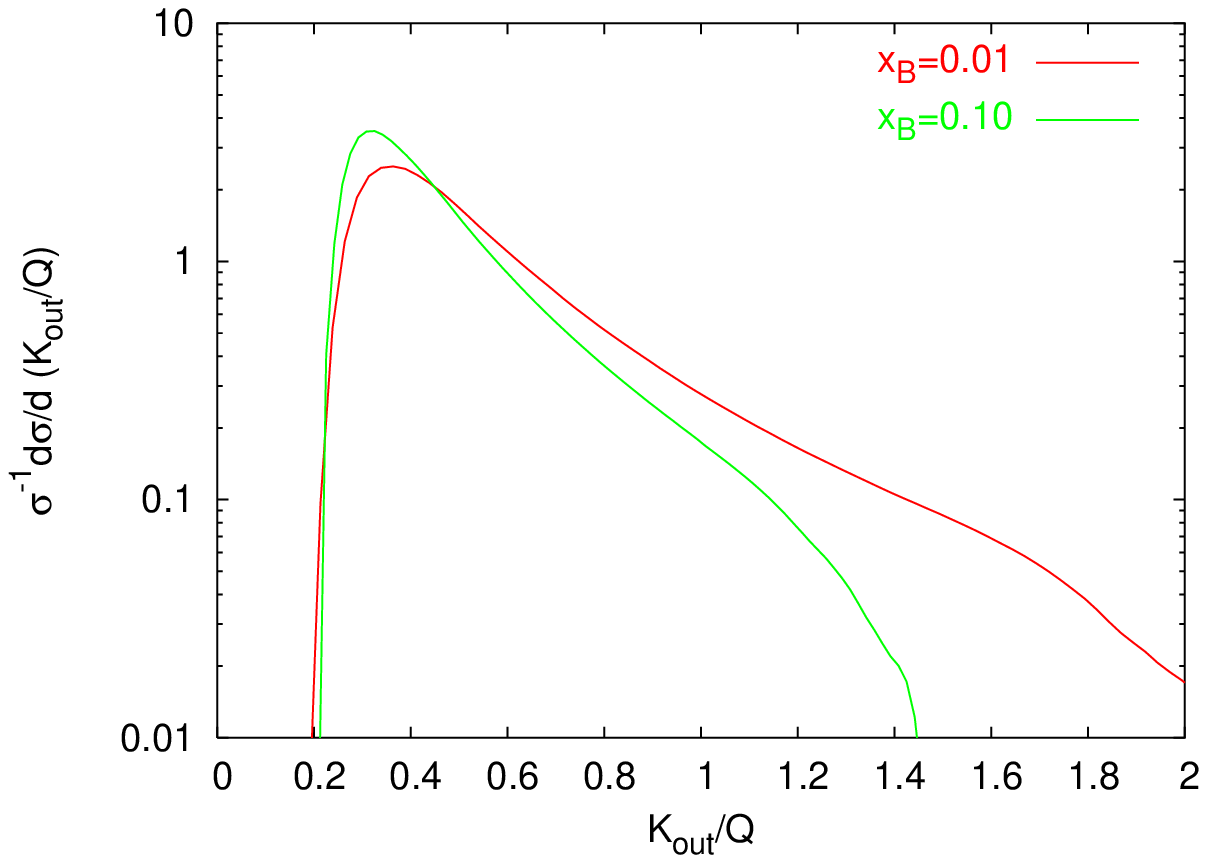,width=0.7\textwidth}
{The full (PT+NP) differential $\Ko$ distribution for two different
values of $\xB$, for $y_-=0.05$, $y_+=2.5$, and
$Q^2=400\GeV^2$.\label{fig:xb=0.1-0.01}}

Finally, Fig.~\ref{fig:xb=0.1-0.01} shows how the distribution depends
on the value of $\xB$. The distribution dies faster by increasing
$\xB$. Indeed, for increasing $\xB$ the centre-of-mass energy of the
hard system decreases, so that the phase space for producing hadrons
with large out-of-plane momentum is reduced.  Moreover, by increasing
$\xB$ the configuration with incoming quark or antiquark ($\conf=2,3$)
becomes more probable.

\section{Discussion and conclusion \label{sec:Discussion}}
In this paper we have extended our knowledge of three-jet physics to
the case of a near-to-planar observable in hard electron-proton
scattering, defined analogously to the thrust-minor in $\ee$
annihilation.  On the PT side, this observable exhibits a rich colour
and geometry dependence. Large angle gluon radiation contributes in
setting the relevant scales for the PT radiator (see
\eqref{eq:QCs}).  The radiator itself is a sum of
three contributions, one for each emitting parton, each one
characterized by a different hard scale: for a quark or antiquark it
is the invariant mass of the fermion system, for a gluon it is its
(invariant) transverse momentum with respect to the fermion pair. Such
a structure, due to the universality of soft radiation, is found to be
common to all near-to-planar $3$-jet shape variables encountered so
far (see \cite{Tmin,Dpar,hh}).  

Since the observable $\Ko$ is uniform in rapidity, hard parton recoil
contributes both to the definition of the event plane and to the
observable. It contributes then to the PT distribution at SL level.

The thrust major axis is determined only by the hard parton system and
cannot be changed (at least at SL accuracy) by secondary soft or
collinear parton radiation. This implies that there exists no
correlation between the up- and down-regions (see the definition of
the event-plane \eqref{eq:ev-plane}).  This property makes the
analysis of our observable much simpler than that of the thrust minor
$T_m$ distribution in $\ee$. In particular, we are able to write a
closed formula for the SL function $\cS_{\conf}$ in \eqref{eq:SL}
which embodies the contributions of hard parton recoil.

Our observable exhibits $1/Q$ power corrections arising from the
running of the coupling into the infrared region.  For $\Ko\gg\LQCD$
they can be taken into account as a shift of the PT distribution (see
\eqref{eq:shift}). The shift $\de \Ko^{(\conf)}$ depends
logarithmically on the observable itself,
so that the effect of the shift is actually a deformation of the PT
distribution.  Since $\Ko$ accumulates contributions from partons of
any rapidity, one has to carefully consider effective rapidity
cut-off.  Along the beam axis the rapidity is bounded by the
experimental resolution $\eta_0$. Along the direction of the two
outgoing hard partons, it is their displacement from the event plane
which provides an effective rapidity cut-off. Averaging such a NP
correction over the hard parton PT recoil distribution gives rise to
contributions of order $\ln\Ko$ or $1/\sqrt{\as}$ according to the
$\Ko$ r\'egimes.  The form of the power corrections can be simply
interpreted on the basis of the event plane kinematics (see
\eqref{eq:ev-plane}), as explained in detail in section~\ref{sec:QCD}.

As in other $3$-jet observables, the magnitude of the NP shift is
expected to be roughly twice as large as the corresponding one for
$2$-jet shape variables (the characteristic weight $2C_F$ of the DL
contribution of the logarithm of $2$-jet distribution becomes here a
$2C_F+C_A$ weight).  This implies that higher order power corrections
may become important, see \cite{KS}. The comparison with data (not
yet available) will be needed to check the validity of our method.

\section*{Acknowledgments}
We started with Yuri Dokshitzer the adventure of analysing multi-jet
observables, this is one of the many applications. We are then
grateful to him for many illuminating discussions and suggestions.  We
thank also Gavin Salam for helpful discussions, suggestions and support in
the numerical analysis.
\newpage
\appendix

\section{Kinematics and elementary partonic cross-sections \label{App:El}}
For the elementary process \eqref{eq:El-proc} we may write the parton
momenta $P_1$, $P_2$ as
\begin{equation}
  \label{eq:parton}
\begin{split}
& P_1 = \frac{Q}{2\ics}(1,0,0,-1) \>,\quad
P_2 = \frac{Q}{2}(z_0,0,T_M,z_3) \>,\quad
\\& 
z_0 = \isp\!+\!\frac{(1\!-\!\isp)(1\!-\!\ics)}{\ics} \>,\quad
z_3 = \isp\!-\!\frac{(1\!-\!\isp)(1\!-\!\ics)}{\ics} \>,
\end{split}
\end{equation}
in terms of the variables in \eqref{eq:El-kin}. Distinguishing $P_2$
and $P_3$ according to \eqref{eq:P2P3}, the variable $\isp$ is given
in terms of $T_M$ by
\begin{equation}
\label{eq:psiT_M}
\isp = \frac{1}{2}\left[1-\sqrt{1-\frac{\ics T_M^2}{1-\ics}}\right]\>,
\end{equation}
and in terms of $y_2$ by the inverse of \eqref{eq:El-y3}, which is
\begin{equation}
\label{eq:ispxy3}
\begin{split}
&\isp=\frac{1-\ics}{2(2\ics-1)}\left[
\sqrt{1+\frac{4\ics(2\ics-1)y_2}{(1-\ics)^2}}-1\right]\>,\quad\left\{
\begin{split}
& 0<\ics<\frac{3}{4},\quad 0<y_2<\frac{1}{4\ics}\\
& \frac{3}{4}<\ics<1,\quad 0<y_2<\frac{2(1-\ics)^2}{\ics(2\ics-1)}
\end{split}\right.\\
&\isp=\frac{\ics^2 y_2-(1-\ics)^2}{(2\ics-1)(1-\ics+\ics y_2)}\>,\qquad
\qquad\qquad\qquad\> \frac{3}{4}<\ics<1,\>
\frac{2(1-\ics)^2}{\ics(2\ics-1)}<y_2<\frac{1-\ics}{\ics}
\end{split}
\end{equation}

Note that at any fixed value of $y_2$, the freedom to vary $x$ results
in events with a range of $T_M$ values. If $y_2<\frac{1}{3}$, the 
kinematic upper limit on $x$
is $x=\frac{1}{1+y_2}$, at which $\isp=\half$ and the lower bound on 
$T_M$ is therefore $\sqrt{y_2}$. The lower limit on $x$ is the Bjorken
variable $\xB$, at which $\isp=\isp(\xB,y_2)<\xB y_2$. (This is obtained
by expanding \eqref{eq:ispxy3} about $\xB=0$: the difference from
equality is of order $\xB^2$.) Consequently, $T_M<2\sqrt{y_2}$.
Therefore, restricting ourselves to events in the range \eqref{eq:yc},
with $y_{-}<\frac{1}{3}<y_+$, leads to a selection of events whose
values of $T_M$ lie in the range
\begin{equation}
\label{eq:TMrange}
\sqrt{y_{-}}<T_M<2\sqrt{y_{+}}\>.
\end{equation}
This of course does not select all events with $T_M$ in this range; it is
merely a means of selecting events whose $T_M$ does not lie outside it,
so that we do not need to consider logarithms of $T_M$. The phase-space
in terms of the variables $(\ics,\isp)$ then becomes:
\begin{equation}
\label{eq:isp_pm}
\begin{split}
&\xB<x<x_M\>,\quad x_M=\left\{\begin{split}
&\frac{1}{1+y_-}\>,\>\>\>y_-<1/3\\
&\frac{1}{4y_-}\>,\>\>\>\>\>\>\>\>y_->1/3
\end{split}\right.\>,\\
&\isp_-<\isp<\isp_+\>,\quad
\isp_-=\isp(x,y_-)\>,\quad
\isp_+=\left\{\begin{split}
&\isp(x,y_+)\>,\>\>\>x<{\textstyle\frac{1}{4y_+}}\\
&\half\>,\qquad\quad\>\>x>{\textstyle\frac{1}{4y_+}}
\end{split}\right. .
\end{split}
\end{equation}

Since we consider only diagrams with the exchange of a single photon or
$Z_0$, the individual partonic cross-sections may be decomposed 
according to:
\begin{equation}
\label{eq:partonicxs}
\begin{split}
\frac{d\hat{\sigma}_{\can,\conf,f}}{d\ics\,d\isp\,dQ^2} = 
\frac{\al^2\as}{Q^4}
& \Bigl\{C^f(Q^2)\left[\left(2-2y+y^2\right)C_T^{\can,\conf}(\ics,\isp)
+2(1-y)C_L^{\can,\conf}(\ics,\isp)\right]\\ 
& +D^f(Q^2)y(2-y)C_3^{\can,\conf}(\ics,\isp)\Bigr\}\>,
\end{split}
\end{equation}
into transverse, longitudinal and parity-violating terms.
Here $y$ is defined by
\begin{equation}
\label{eq:y}
y = \frac{(Pq)}{(PP_e)} = \frac{Q^2}{\xB s} \>,
\end{equation}
where $P_e$ is the momentum of the incident electron, and
$s$ is the centre-of-mass energy squared of the collision (we neglect
the proton mass).
The flavour-dependent functions $C^f(Q^2)$ and $D^f(Q^2)$ show how the
$\gam$ and $Z$ exchange diagrams combine:
\begin{equation}
\label{eq:Z-cross}
\begin{split}
C^f(Q^2) &= e_f^2-\frac{2e_f V_f V_e}{\sin^2 2\theta_W}
\left(\frac{Q^2}{Q^2\!+\!M^2}\right)+\frac{(V_f^2\!+\!A_f^2)(V_e^2\!+\!A_e^2)}
{\sin^4 2\theta_W}\left(\frac{Q^2}{Q^2\!+\!M^2}\right)^2\>,\\
D^f(Q^2) &= -\frac{2e_f A_q A_e}{\sin^2 2\theta_W}
\left(\frac{Q^2}{Q^2\!+\!M^2}\right)+\frac{4V_f A_f V_e A_e}
{\sin^4 2\theta_W}\left(\frac{Q^2}{Q^2\!+\!M^2}\right)^2.
\end{split}
\end{equation}
The functions $C_{T/L/3}^{\can,\conf}(\ics,\isp)$ are the one loop 
QCD elementary square matrix elements for the hard vertex (there is no
contribution of order $\cO{\as^0}$ since we require transverse jets).
We have \cite{dishard}
\begin{equation}
\begin{split}
&C_T^{q,3} = C_F\left[\frac{\ics^2+\isp^2}{(1-\ics)(1-\isp)}
+2(1+\ics\isp)\right]\,,
\hspace{0.4cm} 
C_T^{g,1} = \left[\ics^2+(1-\ics)^2\right]
\frac{\isp^2+(1-\isp)^2}{\isp(1-\isp)}\,,\\
&C_L^{q,3} = C_F\cdot 4\ics\isp\,, 
\hspace{4.8cm} 
C_L^{g,1} = 8\ics(1-\ics)\,, \\
&C_3^{q,3} = C_F\left[\frac{\ics^2+\isp^2}{(1-\ics)(1-\isp)}
+2(\ics+\isp)\right]\,,
\hspace{0.5cm} 
C_3^{g,1} = 0\,,\\
&C_{T/L/3}^{q,2}(\ics,\isp) = C_{T/L/3}^{q,3}(\ics,1\!-\!\isp)\,,\>\>
C_{T/L}^{\bq,\conf}(\ics,\isp) = C_{T/L}^{q,\conf}(\ics,\isp)\,,\>\> 
C_3^{\bq,\conf}(\ics,\isp) = -C_3^{q,\conf}(\ics,\isp)\,.
\end{split}
\end{equation}

\section{Resumming QCD emission \label{App:Radiation}}
The QCD results illustrated in section \ref{sec:QCD} are based
on calculations similar to those performed for other $3$-jet shape
distributions in $\ee$ annihilation \cite{Tmin,Dpar} and in
hadronic collisions \cite{hh}.  
Here we follow the method and use the results developed and obtained
there. We do not report the full details of the calculations but
simply discuss the main features characteristic of the present
distribution: for details refer to the later appendices and to the
previous papers. In particular here we
\begin{itemize}
\item deduce, in the present context, the factorized structure in
  \eqref{eq:Factor} and identify the hard scale entering the incoming
  parton distribution $\cP$;
\item obtain the PT radiation factor at the SL accuracy \eqref{eq:PT-cA};
\item compute the leading order NP corrections giving rise to the shift
  \eqref{eq:shift};
\item show that, for $\Ko$ in the range \eqref{eq:Komin}, the $\eta_0$
  dependence enters only in the NP shift.
\end{itemize}

\subsection{Resummation of the distribution}
Considering the region $\Ko\ll Q$, the starting point is the
factorization of the square amplitude for the emission of $n$
secondary soft partons in the process \eqref{eq:partonproc}
(contributions from hard collinear secondary emission are included
later). We have
\begin{equation}
  \label{eq:Mn1}
  |M_{\can,\conf,f,\,n}(k_1\ldots k_n)|^2
\simeq|M_{\can,\conf,f,\,0}|^2\cdot 
S_{\conf,\,n}(k_1\ldots k_n)\>.
\end{equation}
The first factor is the Born squared amplitude which gives rise to the
elementary hard distribution $d\hat \sigma_{\can,\conf,f}$ in
\eqref{eq:QCD-dsigma}. The second factor is the distribution in the
soft partons emitted from the system of the three hard partons
$p_1,p_2$ and $p_3$ in \eqref{eq:partonproc}.  It depends on the
colour charges of the emitters which are identified by the
configurations $\conf=1,2,3$, the index of the primary gluon momentum.
Since soft radiation is universal and does not change the nature of
incoming parton, $S_{\conf,\,n}$ does not depend on $\can$ and $f$.
For soft emission, the primary partons $p_a$ differ from the hard
primary parton momenta $P_a$ (depending on $Q,\ics,\isp$, see Appendix
\ref{App:El}) by small recoil components.

By using \eqref{eq:Mn1}, the soft contributions to the cross-section
are resummed by
\begin{equation}
\begin{split}
\frac{d\sigma(\Ko,y_{\pm})}{d\xB\,dQ^2} = 
\sum_{\can,\conf,f}\int_{\xB}^{x_M} 
d\ics & \int_{\isp_-}^{\isp_+}d\isp
\left(\frac{d\hat\sigma_{\can,\conf,f}}{d\ics\,d\isp\,dQ^2}\right)
\int_0^1 dx_1 \cP_{\can,f}\left(x_1,\mu\right)\\
&\times\sum_n\,\frac{1}{n!}\,\prod_{i=1}^n\,\int \frac{d^3k_i}{\pi\om_i}\>
\int d\Phi_n\cdot S_{\conf,\,n}\>,
\end{split}
\end{equation}
with $x_M$ given in Appendix \ref{App:El} and $x_1$ the momentum
fraction \eqref{eq:p1} of the incoming parton and $\mu$ the (small)
subtraction scale. The phase space $d\Phi_n$ fixes the observable
$\Ko$, the event plane and $\ics$, the Bjorken variable for the hard
elementary distribution $d\hat\sigma\,$.  The momentum fraction of the
parton entering the hard scattering is then
\begin{equation}
  \label{eq:X1}
\frac{\xB}{\ics}=x_1\prod_{i\in\cC_1}z_i\>,
\end{equation}
where $z_i$ are the collinear splitting fractions for radiation of secondary
partons $k_i$ in the region $\cC_1$ collinear to $p_1$.  The phase
space $d\Phi_n$ is then given by
\begin{equation}
\label{eq:dPhi}
\begin{split}
d\Phi_n=& dp_{2x}\,dp_{3x}\,
\Theta\left(\Ko-|p_{2x}|-|p_{3x}|-{\sum_i}'|k_{ix}|\right)
\de\left(\xB\!-\!\ics x_1\prod_{i\in\cC_1}z_i\right)\\
&\de\left(p_{2x}+{\sum_{i\in U}}'k_{ix}+\frac{1}{2}{\sum_i}'' k_{ix}\right)
 \de\left(p_{3x}+{\sum_{i\in D}}'k_{ix}+\frac{1}{2}{\sum_i}'' k_{ix}\right),
\end{split}
\end{equation}
where the last two delta functions fix the event plane, as shown in
\eqref{eq:ev-plane}.

So the radiation factor $\cI$ in \eqref{eq:QCD-dsigma} takes the form
\begin{equation}
  \label{eq:cI}
\cI_{\can,\conf,f}
= \ics\int_0^1 dx_1\cP_{\can,f}\left(x_1,\mu\right)\>
\sum_n\,\frac{1}{n!}\,\prod_{i=1}^n\,\int \frac{d^3k_i}{\pi\om_i}\>
\int d\Phi_n\cdot S_{\conf,\,n}\,.
\end{equation}
To resum the secondary parton emissions we use the factorization
structure of the soft emission factor $S_{\conf,\,n}$. The phase space
$d\Phi_n$ can be factorized by Mellin and Fourier transforms. We get
\begin{equation}
\label{eq:dPhin}
\begin{split}
&
\int d\Phi_n=
\frac{1}{\xB}\int\frac{dN}{2\pi i}
\left(\frac{\ics x_1}{\xB}\right)^{N-1}
\int\frac{d\nu\,e^{\nu\Ko}}{2\pi i\nu} 
\\&\times 
\prod_{a=2,3} \int_{-\infty}^{\infty}\frac{\nu d\be_a dp_{ax}}{2\pi}
e^{-\nu(|p_{ax}|-i\be_ap_{ax})}
\prod_i \eps(z_i)\,U(k_i)\,,
\end{split}
\end{equation}
where
\begin{equation}
  \label{eq:eps}
\begin{split}  
 \left\{
\begin{split}
& \eps(z)=z^{N-1}\,, \quad \mbox{for}\quad k\in \cC_1\,,\\
& \eps(z)=1\,, \qquad\>\>\> \mbox{for}\quad k\in \!\!\!\!| \>\>\cC_1\,,
\end{split}
\right.
\end{split}
\end{equation}
and 
\begin{equation}
  \label{eq:U}
\begin{split}  
\left\{\>\>
\begin{split}
&U(k)= u_2(k_x)\:\,\equiv e^{-\nu(|k_{x}|-i\be_2 k_{x})}\quad\, 
\mbox{for}\>k_y>0\>,\>\>\eta_k<\eta_0\>,\\
&U(k)= u_3(k_x)\:\,\equiv e^{-\nu(|k_{x}|-i\be_3 k_{x})}\quad\, 
\mbox{for}\>k_y<0\>,\>\>\eta_k<\eta_0\>,\\
&U(k)= u_0(k_x)\:\,\equiv e^{i\frac{\nu}{2}(\be_2+\be_3)k_{x}}
\qquad \mbox{for}\> \eta_k>\eta_0\>.
\end{split}
\right.
\end{split}
\end{equation}
The last source $u_0(k)$ corresponds to partons emitted in the beam
region, which contributes only to the momentum conservation and not to
the observable $\Ko$.  The near-to-planar region $\Ko \ll Q$
corresponds to the region  $\nu Q\gg 1$ in the Mellin variable.

Resummation can be now performed and one obtains
\begin{equation}
  \label{eq:cI-fine}
\begin{split}
\cI_{\can,\conf,f}(\Ko,\ics,\isp,Q,\xB,\eta_0) 
&\!=\!\int\frac{d\nu\,e^{\nu\Ko}}{2\pi i\nu}\!
\int_0^1 \frac{dx_1}{x_1}\int\frac{dN}{2\pi i}
\left(\frac{\ics\,x_1}{\xB}\right)^{N}\cP_{\can,f}(x_1,\mu)\\
&\times \prod_{a=2,3}\int\frac{\nu d\be_a dp_{ax}}{2\pi} 
e^{-\nu(|p_{ax}|-i\be_ap_{ax})}\> e^{-\cE_{\conf}}\>,
\end{split}
\end{equation}
where $\cE_{\conf}$ is the full radiator for the secondary emission
and is given by
\begin{equation}
  \label{eq:cR}
  \cE_{\conf} =
\int \frac{d^3k}{\pi \om}W_{\conf}(k)\,\left[1-U(k)\eps(z)\right]\>.
\end{equation}
The $1$ in the brackets here represents virtual gluon emission. The
radiator $\cE_\conf$ depends on the source variables, $\nu,\be_a,N,\eta_0$, 
on the primary parton momenta $p_a$, which are given in terms of
$Q,\ics,\isp$ and the recoil components $p_{ax}$ (the other components
can be neglected in the soft limit), and on the subtraction scale
$\mu$ needed to regularize the collinear divergence.  Here
$W_{\conf}(k)$ is the two-loop distribution for the emission of the
soft gluon from the three hard partons 
 $p_a$
in the configuration $\conf$:

\begin{equation}
W_{\conf}(k) = \frac{N_c}{2}\left(w_{\conf a}(k)+w_{\conf b}(k)-\frac{1}{N_c^2}w_{ab}(k)\right)\>,\qquad a\neq b\neq \conf\>,
\end{equation}
where $w_{ab}(k)$ is the standard distribution for emission of a soft
gluon from the $ab$-dipole
\begin{equation}
w_{ab}(k)=\frac{\as(k_{ab,t})}{\pi k_{ab,t}^2}\>,\qquad
k_{ab,t}^2=\frac{2(p_a k)(k p_b)}{(p_a p_b)}\>,
\end{equation}
and $\as$ is in the physical scheme \cite{CMW}.

\subsection{Factorization of incoming parton distribution}
In the present formulation, the factorization \eqref{eq:Factor} of the
distribution $\cI$ results by splitting the source
\begin{equation}
  \label{eq:split}
  [1-U(k)\eps(z)]=[1-U(k)]\>+\>[1-\eps(z)]\,U(k)\>,
\end{equation}
so that the radiator $\cE$ can be expressed as the sum of two terms (we
write explicitly only the $N$ and $\mu$ dependence)
\begin{equation}
  \label{eq:RG}
\cE_{\conf}(N,\mu)=\cR_{\conf}\>+\>\Gamma_{\conf}(N,\mu)\>,
\end{equation}
with
\begin{equation}
  \label{eq:R+G}
\begin{split}
\cR_{\conf}=&\int\frac{d^3k}{\pi\om}\>W_{\conf}(k)\>[1-U(k)]\>,\\
\Gamma_{\conf}(N,\mu)=&\int_{\cC_{1}}\frac{d^3k}{\pi\om}\>
W_{\conf}(k)\> \left[1-z^{N-1}\right]\,U(k)\>. 
\end{split}
\end{equation}
These two terms will be evaluated for small $\Ko\sim\nu^{-1}$.

Consider first the collinear singular piece $\Gamma_{\conf}(N,\mu)$.
Here the source $U(k)$ provides an upper bound for the integration
frequencies of order $ \nu^{-1}\sim\Ko$. This is shown by using, in
the integral \eqref{eq:R+G},
\begin{equation}
  \label{eq:rho}
[1-u_a(k_x)]\>\simeq\>
\Theta\left(\,|k_x|-\rho_a^{-1}\,\right)\>,\qquad 
\rho_a=e^{\gam_E}\nu\sqrt{1+\be_a^2}\>,
\end{equation}
for $a=2,3$,
which is valid for large $\nu$ within SL accuracy (see for instance
Appendix~C of Ref.~\cite{Dpar}).  Using this approximate expression
one also shows \cite{hh} that $\Gamma_{\conf}(N,\nu)$ does not
depend on $\eta_0$, as long as $\Ko$ is taken in the range
\eqref{eq:Komin}.  As shown in \cite{hh}, $\Gamma(N,\nu)$
is the soft part of the anomalous dimension that evolves the parton
distributions $\cP$ from the small subtraction scale to the scale
$\rho_a^{-1}$.  It is accurate at two loop order provided we use the coupling
in the physical scheme \cite{CMW}, and it is diagonal in the
configuration index since soft radiation is universal and does not
change the nature of the incoming parton. But in general one needs the
full anomalous dimension (including the hard non diagonal pieces) so
that the incoming partons may change from a quark to a gluon and vice
versa. The resulting exponent $\Gamma$ becomes a matrix. Upon
integration over $x_1$ and the Mellin variable $N$
\begin{equation}
  \label{eq:cPhard}
\int_0^1\frac{dx_1}{x_1}\int\frac{dN}{2\pi i} 
\left(\frac{\ics\,x_1}{\xB}\right)^{N}\!
\cP(x_1,\mu)\cdot e^{-\Gamma(N,\mu)}
\,\simeq\, \cP\left(\frac{\xB}{\ics},\Ko\right)\>,
\end{equation}
one obtains the parton distribution $\cP(\Ko)$ evolved from the
subtraction to the hard scale $\Ko$, up to terms that are beyond
SL\footnote{The more accurate expression
$\sqrt{\cP(\rho_2^{-1})\cP(\rho_3^{-1})}$ is required for some of the
NP contributions.}. The $\mu$ dependence in
$\cP(\mu)$ and $\Gamma(N,\mu)$ is cancelled.
We do not consider here power corrections since they are of second
order ($1/Q^2$), see \cite{DMW}.

We consider then the piece $\cR_{\conf}$. This is a CIS quantity
($[1\!-\!U(k)]\!\to\!0$ for $k_t\!\to\!0$) which produces the
radiation factor $\cA_\conf$ in \eqref{eq:Factor}
\begin{equation}
\label{eq:cA}
\cA_{\conf}(\Ko,\ics,\isp,Q,\eta_0)=\int\frac{d\nu\,e^{\nu\Ko}}{2\pi i \nu}
\,\hat{\cA}_{\conf}(\nu,\ics,\isp,Q,\eta_0)\>,
\end{equation}
with the Mellin moment
\begin{equation}
  \label{eq:hat-cA}
\hat{\cA}_{\conf}=\int \prod_{a=2,3} \frac{\nu d\be_a dp_{ax}}{2\pi}\,
e^{-\nu (|p_{ax}|-i\be_ap_{ax})}\>
e^{-\cR_{\conf}}\>.
\end{equation}
All the PT contributions are finite and they can be evaluated to SL
accuracy by using the approximation \eqref{eq:rho} for the
sources. The PT resummed expression $\cR^{\PT}_{\conf}$ is then obtained
by evaluating the integral \eqref{eq:R+G} of $\cR_{\conf}$ using
\eqref{eq:rho}. 

However the virtual momentum $k_t$, entering the SL resummed running
coupling in $\cR_{\conf}$, cannot be prevented from going to zero. Although
the contribution from this NP region is (formally) highly subleading
(power suppressed), it is phenomenologically quite important for $\Ko$
not too large \cite{NPstandard}.  For $k_t$ very small (in particular
for $|k_x|\ll\nu^{-1}\sim\Ko$) the approximation \eqref{eq:rho} is no
longer valid and we have instead
\begin{equation}
\label{eq:smallU}
[1-u_a(k)]\simeq \nu\,(|k_x|-i\be_ak_x)\>,
\end{equation}
so that the leading NP correction to $\cR_{\conf}$ is 
\begin{equation}
\label{eq:dKo0}
\de \cR_{\conf}\simeq \nu\cdot \de \Ko^{(\conf)}\>,
\end{equation}
corresponding to a shift in $\Ko$ in \eqref{eq:cA}. To evaluate this NP
shift we use the dispersive method \cite{DMW} to represent the running
coupling and we evaluate the leading power-suppressed contribution
from \eqref{eq:smallU}.

The two contributions to the radiator 
\begin{equation}
  \label{eq:Radtot}
  \cR_{\conf}=  \cR^{\PT}_{\conf}+ \de \cR_{\conf}\>,
\end{equation}
together with the expression for the radiation factor $\cA_{\conf}$ will be
described in the next two subsections.  

\subsection{PT Radiation Factor}
The PT radiator $\cR_{\conf}^{\PT}$ to SL accuracy coincides with a
particular case of the PT radiator computed for $T_m$ in $\ee$
annihilation (where we set $\gamma=\be_1=0$ in eq.~(4.8) of Ref.~\cite{Tmin}). 
For $\Ko$ in the range \eqref{eq:Komin} this contribution is $\eta_0$ 
independent, see \cite{hh}. The explicit calculation is given in 
Appendix \ref{App:RadPT}, and the result may be conveniently separated
into two terms:
\begin{equation}
\label{eq:Rad-PT}
\cR_{\conf}^{\PT}= R_{12}^{(\conf)}(\rho_2)+R_{13}^{(\conf)}(\rho_3)\>,
\end{equation}
where
\begin{equation}
\label{eq:R1a}
R_{1a}^{(\conf)}(\rho_a)=
\half C_1^{(\conf)} r(\rho_a,\zeta_1^{(\conf)}Q_1^{(\conf)})+
C_a^{(\conf)} r(\rho_a,\zeta_a^{(\conf)}Q_a^{(\conf)})\>,
\end{equation}
with the function $r$ given in \eqref{eq:r} and colour charges
$C_{a}^{(\conf)}$ and hard scales $Q_{a}^{(\conf)}$ given in
\eqref{eq:QCs}.  The term $R_{12}^{(\conf)}$ corresponds to the
emission in the up-region off parton $p_1$ and $p_2$. Similarly for
$R_{13}^{(\conf)}$. For the PT evaluation we can expand to SL accuracy
\begin{equation}
\label{eq:Rad-SL}
\begin{split}
\cR_{\conf}^{\PT}
=R_{\conf}\left(\nu,\ics,\isp,Q\right)
+\left(C_T\gam_E +C_{12}^{(\conf)}\ln\sqrt{1\!+\!\be_2^2}
+C_{13}^{(\conf)}\ln\sqrt{1\!+\!\be_3^2}\right)r'(\nu,Q)\,,
\end{split}
\end{equation}
with $R_{\conf}$ the DL exponent function introduced in
\eqref{eq:PT-rad} and $r'$ the SL function \eqref{eq:r'}.  Since the
radiator is independent of the recoils we can integrate over $p_{ax}$
and, using the expansion \eqref{eq:Rad-SL}, we obtain the Mellin moment
\begin{equation}
  \label{eq:PT-hatcA}
\begin{split}
\hat{\cA}^{\PT}_{\conf}&(\nu,\ics,\isp,Q)
=\prod_{a=2,3}\int_{-\infty}^{\infty}\frac{d\be_a}{\pi(1+\be_a^2)}
e^{-R^{(\conf)}_{1a}} \\
&\simeq e^{-R_{\conf}\left(\nu,\ics,\isp,Q\right)}\,
e^{-\gam_E C_T\,r'}\cF\left(C_{12}^{(\conf)}\,r'\right)
\cF\left(C_{13}^{(\conf)}\,r'\right),
\qquad r'=r'(\nu,Q)\>,
\end{split}
\end{equation}
with $\cF$ given in \eqref{eq:cF}. 

The PT radiation factor is obtained by integrating over the Mellin 
variable $\nu$. We make use of the operator identity
\begin{equation}
\label{eq:op-identity}
\int\frac{d\nu\,e^{\nu\Ko}}{2\pi i\nu}\>G(\nu) = \frac{1}{\Gamma
\left(1+\frac{\partial}{\partial\ln\Ko}\right)}\>G(\Ko^{-1})
\end{equation}
for any logarithmically varying function $G$. (To prove this,
multiply both sides by the $\Gamma$-function operator and use
the definition $\Gamma(z) = \int_0^\infty dx\,x^{z-1}e^{-x}$.)
Thus to SL accuracy we have
\begin{equation}
  \label{eq:PT-cA'}
\cA^{\PT}_\conf(\Ko,\ics,\isp,Q) = 
\frac{\hat{\cA}^{\PT}_\conf(\Ko^{-1},\ics,\isp,Q)}
{\Gamma\left(1+C_T r'(\Ko^{-1},Q)\right)}\>,
\end{equation}
which corresponds to the PT result in \eqref{eq:PT-cA}, \eqref{eq:r}
and \eqref{eq:SL}.

\subsection{Radiation Factor including NP corrections}
The NP correction to the radiator is computed in Appendix
\ref{App:RadNP} following the standard procedure (see \cite{Tmin}):
we assume that the running coupling can be defined even at small
momenta via a dispersion relation \cite{DMW}, and we take into account 
the non-inclusiveness of the observable via the Milan factor. The NP
correction is proportional to the parameter $\cp$, given in
\eqref{eq:cp}, expressed in terms of the integral of the running 
coupling in the small momentum region. This parameter takes into
account merging of NP and PT corrections at two loops and is the same
as entering all the jet-shape distributions so far studied. We obtain
\begin{equation}
\label{eq:RadNP}
\delta \cR_{\conf}= \nu\cp\left(
C_1^{(\conf)}\ln\frac{\ics e^{\eta_0}Q_1^{(\conf)}}{Q}+
\sum_{a=2,3}C_a^{(\conf)}\ln\frac{\zeta Q_a^{(\conf)}}{|p_{ax}|}
\right)\>,\qquad \zeta = 2e^{-2}\>,
\end{equation}
where $Q_a^{(\conf)}$ are the scales introduced in \eqref{eq:QCs}.
The NP radiator is made up of three contributions, one for each emitting
parton.  The hard scales are determined by large angle NP gluons; they
are therefore the same as in the PT case. What makes the difference
between the three contributions is the NP gluon rapidity
cutoff. Actually, when a NP gluon is emitted from the incoming parton
$p_1$, its rapidity is bounded by the experimental resolution
$\eta_0$. This is not the case for the emission from $p_2$ or $p_3$,
where it is the out-of-plane momentum of the recoiling parton which
provides an effective rapidity cutoff.

The first order NP correction to the Mellin moment $\hat\cA_{\conf}$ in
\eqref{eq:hat-cA} is given by
\begin{equation}
  \label{eq:sigmahat}
\hat\cA_\conf(\nu) = \prod_{a=2,3}\int
\frac{\nu d\be_a dp_{3x}}{2\pi}\,e^{-\nu(|p_{ax}| - i\be_a p_{ax})}\>
e^{-R^{\PT}_{\conf}}\,\left\{1-\de \cR_{\conf}\right\}\>,
\end{equation}
where $R^{\PT}_{\conf}$ is the PT radiator in \eqref{eq:Rad-PT}. The
first term yields the PT contribution in \eqref{eq:PT-cA} and so we can
write
\begin{equation}
  \label{eq:sigmahat1}
  \hat \cA_{\conf}(\nu)=  \hat \cA_{\conf}^{\PT}(\nu)-
\nu\,\cp\>f_{\conf}(\nu)\>.
\end{equation}
After the $p_{ax}$ integration the NP correction is given by
\begin{equation}
  \label{eq:f}
f_{\conf}(\nu)\!=\!\prod_{a=2,3}\!\int_{-\infty}^{\infty}\!\!
\frac{d\be_a} {\pi(1\!+\!\be_a^2)}\,e^{-\cR_{\conf}^{\PT}}\!\!
\left(\!C_1^{(\conf)}\ln\frac{\ics e^{\eta_0}Q_1^{(\conf)}}{Q}
\!+\!\sum_{a=2,3}\!C_a^{(\conf)}
[\ln\zeta\,\bnu\,Q_a^{(\conf)}\!+\!\chi(\be_a)]\!\right), 
\end{equation}
with $\chi(\be)=\ln\sqrt{1+\be^2}+\be\tan^{-1}\be$. This function is
growing as $\pi|\be|/2$ at large $\be$. As a consequence, the $\be_a$
integration in \eqref{eq:f} is no longer fastly convergent and we
cannot expand the PT radiator as in \eqref{eq:Rad-SL} but we need to
keep  its exact expression. The region of large $|\be_a|$, which
corresponds to $|p_{ax}|\ll \Ko$, gives the leading NP correction.

The radiation factor $\cA$ is obtained by performing the $\nu$
integration so that
\begin{equation}
  \label{eq:desigma}
  \cA_{\conf}(\Ko)=\cA_{\conf}^{\PT}(\Ko)+\de\cA_{\conf}(\Ko)\>.
\end{equation}
In Appendix~\ref{App:NPdist}, see also \cite{broad,Tmin}, we show that
$\de\cA_{\conf}$ can be expressed (to first order) as a shift (see
\eqref{eq:shift})
\begin{equation}
  \label{eq:ma}
  \de\cA_{\conf}(\Ko)=-\cp \partial_{\Ko}
\int\frac{d\nu\, e^{\nu\Ko}}{2\pi i \nu}\,f_{\conf}(\nu)
\simeq -\de\Ko^{(\conf)}\cdot \partial_{\Ko}\cA_\conf^{\PT}(\Ko)\>,
\end{equation}
where the shift is given by 
\begin{equation}
\label{eq:NPshift}
\begin{split}
&\frac{\de\Ko^{(\conf)}}{\cp}=
C_1^{(\conf)}\ln\frac{\ics e^{\eta_0}Q_1^{(\conf)}}{Q}
\!+\!\sum_{a=2,3}C_a^{(\conf)}\!\left[
\ln\frac{Q_a^{(\conf,\NP)}}{2\bKo}+
\frac{C_T\!-\!C_{1a}^{(\conf)}}{C_T}
\left\{ E_a^{(\conf)}(\bKo^{-1})\!+\!\cC_a^{(\conf)}\right\} \right].
\end{split}
\end{equation}
Here the NP hard scale is defined by
\begin{equation}
\begin{split}
\ln\frac{Q_a^{(\conf,\NP)}}{2\bKo}
&=\ln\frac{\zeta Q_a^{(\conf)}}{\bKo}+
\psi(1+C_T\,r')+\half\psi\left(1\!+\!\frac{C_{1a}^{(\conf)}\,r'}{2}\right)
-\half\psi\left(\frac{1\!+\!C_{1a}^{(\conf)}\,r'}{2}\right)\\
&\simeq
\ln\frac{2\,\zeta Q_a^{(\conf)}}{\Ko}+\cO{r'}\>,
\quad \bKo=e^{-\gam_E}\Ko\>,
\end{split}
\end{equation}
where $r'=r'(\Ko^{-1},Q)$ and the choice of factors of 2 and $e^{-\gam_E}$
is so as to give the simple form for equation \eqref{eq:Laexp2a} below. 
The function $E_a^{(\conf)}$ is given by
\begin{equation}
\label{eq:Ea}
\begin{split}
E_a^{(\conf)}(\bKo^{-1})=\int_{\bKo^{-1}}^{\infty}\frac{d\rho}{\rho}
e^{-R_{1a}^{(\conf)}(\rho) +R^{(\conf)}_{1a}(\bKo^{-1})}
\left(\frac{\cP(\rho_a^{-1})}{\cP(\Ko)}\right)^{\frac12}\>, 
\end{split}
\end{equation}
where the appearence of the parton distributions comes from the
mismatch between the correct scales $\rho_a^{-1}$ and the SL
approximation $\Ko$ in \eqref{eq:cPhard}. They introduce a dependence
on $\tau,\>f$ and $\xB$ which we suppress.

As explained in detail in \cite{Tmin}, these functions result from
an interplay between PT and NP contributions. They can be expressed as
the average of the rapidity length over the Sudakov factor for
emitting in the up- or down-region for $a=2$ or $a=3$ respectively.
Finally,
\begin{equation}
\cC_a^{(\conf)}=
\left(E_a^{(\conf)}(\bKo^{-1})-\frac{1}{C_{1a}^{(\conf)}r'}\right)
\left(\frac{1}{\cF\left(C_{1a}^{(\conf)}r'\right)}-1\right)
\end{equation}
is a regular function of $r'$.

The shift can be expressed as a sum of partial shifts as in
\eqref{eq:shift-tot} with the rapidity integrals given by
\begin{equation}
\label{eq:Ys}
\begin{split}
&Y_{21}^{(\conf)}=Y_{31}^{(\conf)}=
\ln\frac{\ics e^{\eta_0}Q_1^{(\conf)}}{Q}\,,\\
&Y_{22}^{(\conf)}=\ln\frac{Q_2^{(\conf,\NP)}}{2\bKo}\,,\qquad\qquad\>\>\>\>
 Y_{33}^{(\conf)}=\ln\frac{Q_3^{(\conf,\NP)}}{2\bKo}\,,\\
&Y_{23}^{(\conf)}=L_{3}^{(\conf)}(\bKo)+\cC_{3}^{(\conf)}\,,\qquad
 Y_{32}^{(\conf)}=L_{2}^{(\conf)}(\bKo)+\cC_{2}^{(\conf)}\>,
\end{split}
\end{equation}
where we have introduced the NP functions $L_a$ defined as 
\begin{equation}
\label{eq:La3}
\begin{split}
L^{(\conf)}_a(\bKo)=\ln\frac{Q_a^{(\conf,\NP)}}{2\bKo}
+E^{(\conf)}_a(\bKo^{-1})\>.
\end{split}
\end{equation}
These functions have the following asymptotic expansions: from
\eqref{eq:regione1a} one has
\begin{equation}
\label{eq:Laexp1a}
L^{(\conf)}_a(\bKo)=\ln\frac{Q_a^{(\conf,\NP)}}{2\bKo}
\left\{1+\cO{\frac{1}{\as\ln^2\frac{Q}{\Ko}}}\right\}\>,
\qquad\as\ln^2\frac{Q}{\Ko}\gg 1\>,
\end{equation}
while from \eqref{eq:regione2a} one obtains the behaviour of $L_a$ in
the region $\as\ln^2 Q/\Ko\ll 1$:
\begin{equation}
\label{eq:Laexp2a}
\begin{split}
L^{(\conf)}_a(\bKo)=
\frac{\pi}{2\sqrt{C^{(\conf)}_{1a}\,\as}}&-\frac{1}{C^{(\conf)}_{1a}}
\left(\!\half C^{(\conf)}_1\ln\frac{Q_1^{(\conf,\PT)}}{Q_a^{(\conf,\NP)}}+
C^{(\conf)}_a\ln\frac{Q_a^{(\conf,\PT)}}{Q_a^{(\conf,\NP)}}\right.\\
&\left. +\frac{\pi}{4\as}\frac{\partial\ln\cP(\Ko)}{\partial\ln\Ko}
+\frac{\be_0}{6}\!\right)
\!+\!\cO{\sqrt{\as}},
\end{split}
\end{equation}
where the PT hard scale is 
$Q_a^{(\conf,\PT)}=\zeta_a^{(\conf)}Q_a^{(\conf)}$.

\section{The PT radiator \label{App:RadPT}}
The PT radiator is given, to SL accuracy, in terms of $ab$-dipole
radiators
\begin{equation}
r_{ab}(\nu,\be_2,\be_3) = r_{ab}^U(\nu,\be_2)+r_{ab}^D(\nu,\be_3)\>,
\end{equation}
where the `up' and `down' hemisphere components are given by
\begin{equation}
  \label{eq:rab}
\begin{split}
r_{ab}^U(\nu,\be_2)&=\int_{k_y>0}\frac{d^3k}{\pi\om}w_{ab}(k)\,
\left[1-e^{-\nu(|k_x|-i\be_2 k_x)}\right],
\qquad w_{ab}(k)=\frac{\as(k^2_{ab,t})}{\pi k^2_{ab,t}}\>,\\
r_{ab}^D(\nu,\be_3)&=\int_{k_y<0}\frac{d^3k}{\pi\om}w_{ab}(k)\,
\left[1-e^{-\nu(|k_x|-i\be_3 k_x)}\right],
\end{split}
\end{equation}
and $k_{ab,\,t}$ is the invariant transverse momentum of $k$ with
respect to the $P_a,P_b$ hard partons in \eqref{eq:parton}. For the
configuration $\conf=1$, for instance, we have
\begin{equation}
  \label{eq:R12}
R^{\PT}_{1}(\nu,\be_a)=\frac{N_c}{2}\left(r_{12}(\nu,\be_a)
+r_{13}(\nu,\be_a)-\frac{1}{N_c^2}\,r_{23}(\nu,\be_a)\right).
\end{equation}
To evaluate the $ab$-dipole radiator $r_{ab}(\nu,\be_a)$ we work in the
centre of mass system of the $ab$-dipole. We neglect the rapidity cut
\eqref{eq:rapidity}: the correction is beyond our accuracy \cite{hh}.
Denoting by $P^*_a,P^*_b$
and $k^*$ the momenta in this system, we introduce the Sudakov
decomposition
\begin{equation}
  \label{eq:Sud}
     P_a^*=\frac{Q_{ab}}{2}(1,0,0,1)\>,\quad 
     P_b^*=\frac{Q_{ab}}{2}(1,0,0,-1)\>,
\qquad k^*=\al P_a^*+\be P_b^*+\ka\>, 
\end{equation}
where the scales $Q_{ab}^2=2(P_aP_b)$ are given in \eqref{eq:Q_ab}.
Here the two-dimensional vector $\vka$ is the transverse momentum
orthogonal to the $ab$-dipole momenta ($\ka^2=k^2_{ab,t}$). We have
then
\begin{equation}
  \label{eq:ab-cm}
w_{ab}(k)=\frac{\as(\ka^2)}{\pi \ka^2}\>, \qquad
\frac{d^3k}{\pi\omega}=\frac{d^2\ka}{\pi}\frac{d\al}{\al}\>, \quad 
\al>\frac{\ka^2}{Q^2_{ab}}\>.
\end{equation}

Since, neglecting the recoils, the outgoing momenta $P_2$ and $P_3$ are
in the $yz$-plane, the Lorentz transformation is in the $yz$-plane and
our observable $k_x$ remains unchanged. However, the boundary between the
`up' and `down' hemispheres becomes non-trivial, and different for each
dipole, so we must take each in turn.

The $12$-radiator with 
\begin{equation}
k^*=\al P_2^*+\be P_1^*+\ka\>, 
\end{equation}
has up and down hemispheres given by
\begin{equation}
U\>: \ka_y > -\frac{QT_M}{2}\al\>,\qquad
D\>: \ka_y < -\frac{QT_M}{2}\al\>,
\end{equation}
with $T_M$ related to $\isp$ by \eqref{eq:psiT_M}.
The up component is then
\begin{equation}
r_{12}^U(\nu,\be_2)=\int_0^{Q^2_{12}}\frac{d^2\ka}{\pi\ka^2}
\frac{\as(\ka)}{\pi}\int_{\ka^2/Q_{12}^2}^1\frac{d\al}{\al}
\left[1-e^{-\nu(|\ka_x|-i\be_2\ka_x)}\right]
\Theta\left(\al+\frac{2\ka_y}{QT_M}\right)\>,
\end{equation}
Integrating over $\al$ and $\ka_y$ we have
\begin{equation}
r_{12}^U(\nu,\be_2)=\int_{-Q_{12}}^{Q_{12}}\frac{dk_x}{|k_x|}
\left[1-e^{-\nu(|k_x|-i\be_2 k_x)}\right]\cdot\frac{\as(2|k_x|)}{\pi}
\left(\ln\frac{Q_{12}}{2|k_x|}+\frac{1}{2}\ln\frac{QT_M}{2|k_x|}\right)\>.
\end{equation}
To show this we introduced $\ka_y=t\cdot |k_x|$ and used
\begin{equation}
\int_{-\infty}^{\infty}\frac{dt}{\pi(1+t^2)}\as\left(|k_x|\sqrt{1+t^2}\right)
\ln \frac{Q_{ab}}{|k_x|\sqrt{1+t^2}}\simeq
\as(2|k_x|)\ln\frac{Q_{ab}}{2|k_x|}\>.
\end{equation}
We extended the $t$-integration to infinity since it is
convergent, then we integrated over $t$ by expanding $\as$ to
second order. Corrections are beyond SL accuracy.

Finally, using \eqref{eq:rho}, we obtain
\begin{equation}
r_{12}^U(\nu,\be_2)=2\int_{\rho_2^{-1}}^{Q_{12}}
\frac{dk_x}{k_x}\frac{\as(2k_x)}{\pi}\>
\left(\ln\frac{Q_{12}}{2k_x}+\frac{1}{2}\ln\frac{QT_M}{2k_x}\right)\>.
\end{equation}
Similarly the `down' hemisphere contribution is
\begin{equation}
r_{12}^D(\nu,\be_3)=2\int_{\rho_3^{-1}}^{Q_{12}}
\frac{dk_x}{k_x}\frac{\as(2k_x)}{\pi}\>
\left(\ln\frac{Q_{12}}{2k_x}-\frac{1}{2}\ln\frac{QT_M}{2k_x}\right)\>.
\end{equation}

The $13$ radiator has analogous results
\begin{equation}
\begin{split}
r_{13}^U(\nu,\be_2) &= 2\int_{\rho_2^{-1}}^{Q_{13}}
\frac{dk_x}{k_x}\frac{\as(2k_x)}{\pi}\>
\left(\ln\frac{Q_{13}}{2k_x}-\frac{1}{2}\ln\frac{QT_M}{2k_x}\right)
\>,\\
r_{13}^D(\nu,\be_3) &= 2\int_{\rho_3^{-1}}^{Q_{13}}
\frac{dk_x}{k_x}\frac{\as(2k_x)}{\pi}\>
\left(\ln\frac{Q_{13}}{2k_x}+\frac{1}{2}\ln\frac{QT_M}{2k_x}\right)\>.
\end{split}
\end{equation}

The $23$-radiator has a slightly different structure: with 
\begin{equation}
k^*=\al P_2^*+\be P_3^*+\ka\>, 
\end{equation}
the up and down hemispheres are given by
\begin{equation}
U\>: \ka_y > \frac{QT_M}{2}\frac{\be-\al}{1-2\isp}\>,\qquad
D\>: \ka_y < \frac{QT_M}{2}\frac{\be-\al}{1-2\isp}\>.
\end{equation}
The up component is then
\begin{equation}
\begin{split}
r_{23}^U(\nu,\be_2)=\int_0^{Q^2_{23}}&\frac{d^2\ka}{\pi\ka^2}
\frac{\as(\ka)}{\pi}\int_{\ka^2/Q_{23}^2}^1\frac{d\al}{\al}
\left[1-e^{-\nu(|\ka_x|-i\be_2\ka_x)}\right]\\
&\Theta\left(\al-\frac{\ka_y(1-2\isp)}{QT_M}
-\sqrt{\frac{\ka_y^2}{Q^2 T_M^2}+\frac{\ka_x^2}{Q_{23}^2}}\right)\>,
\end{split}
\end{equation}
Performing the integrals as above we obtain
\begin{equation}
r_{23}^U(\nu,\be_2)=2\int_{\rho_2^{-1}}^{Q_{23}}
\frac{dk_x}{k_x}\frac{\as(2k_x)}{\pi}\ln\frac{Q_{23}}{2k_x}\>,
\end{equation}
and similarly
\begin{equation}
r_{23}^D(\nu,\be_3)=2\int_{\rho_3^{-1}}^{Q_{23}}
\frac{dk_x}{k_x}\frac{\as(2k_x)}{\pi}\ln\frac{Q_{23}}{2k_x}\>.
\end{equation}

Assembling the various dipole contributions and including hard
collinear splittings then yields, to SL accuracy,
\begin{equation}
\begin{split}
R_\conf^{\PT}(\nu,\be_2,\be_3) = \frac{1}{2}&\sum_{a=1}^3 C_a^{(\conf)}
\left[r(\rho_2^{},\zeta_a^{(\conf)}Q_a^{(\conf)})+
r(\rho_3^{},\zeta_a^{(\conf)}Q_a^{(\conf)})\right]\\
&+\frac{1}{2}\left(C_2^{(\conf)}-C_3^{(\conf)}\right)
\left[r(\rho_2,QT_M)-r(\rho_3,QT_M)\right]\>.
\end{split}
\end{equation}
Finally, we note that the terms on the second line contribute only at
SL level, since the DL terms cancel. We may therefore change at will the
hard scales in these terms, yielding the result \eqref{eq:Rad-PT}.

\section{NP corrections to the radiator \label{App:RadNP}}
We consider the NP correction $\de r_{ab}$ to the $ab$-dipole
radiator.  In this case, as we shall see, we need to retain both the
recoil momenta $p_{2x}$ and $p_{3x}$ and the rapidity cut $\eta_0$ . We
write the integral in the $ab$-dipole centre of mass variables
$\al,\be$ and $\vka$ introduced in \eqref{eq:Sud} and, to obtain the
NP correction $\de r_{ab}$, we perform the following standard
operations:
\begin{itemize}
\item the running coupling, reconstructed by two loop emission, is
  represented by the dispersive form \cite{DMW}.  Then, the
  $ab$-dipole radiation $w_{ab}(k)$ is written in the $ab$-centre of
  mass system (see \eqref{eq:ab-cm}) in the form
  \begin{equation}
    w_{ab}(k)=\frac{\as(\ka)}{\pi \ka^2}=\int_0^{\infty} 
    \frac{dm^2\,\aef(m)}{\pi(\ka^2+m^2)^2}\>;
  \end{equation}
\item to take into account the emission of soft partons at two loop
  order \cite{Milan}, we need to extend the source $u(k_x)$ to include
  the mass $m$ of the soft system.  We assume
  $k_x=\ka\cos\phi\to\sqrt{\ka^2+m^2}\cos\phi$, with $\phi$ the
  azimuthal angle of $\vka$. Similarly we introduce the mass in the
  kinematical relations such as $\al\be=(\ka^2+m^2)/Q_{ab}^2$ for the
  $ab$-dipole variables;
\item we take the NP part $\de\aef(m)$ of the effective coupling.
  Since it has support only for small $m$, we take the leading part
  of the integrand for small $\ka$, and $m$.  In particular we
  linearize the source $U(k)$
  \begin{equation}
\label{eq:linearU}
    \left[1-U(k)\right]\>\to\> \nu\sqrt{\ka^2+m^2}\>|\cos\phi|\>
\Theta(\eta_0-\eta_k)\>.
\end{equation}
Recall that $\eta_k$ is the rapidity of $k$ in the Breit frame
\eqref{eq:qP}. Here we have neglected terms proportional to $\be_2$
and $\be_3$ since they vanish, by symmetry, upon the $\be_a$
integration;
\item the recoil component $p_{ax}$ of the outgoing parton $p_a$ does
  provide an effective cut in the soft gluon rapidity along the
  outgoing parton \cite{Tmin,broad}. This is due to a real-virtual
  cancellation which takes place when the angle of the outgoing parton
  $p_a$ with the event plane exceeds the corresponding angle of the
  soft gluon. The detailed analysis of real and virtual pieces entails
  that the contribution from the observable
  $\sqrt{\ka^2+m^2}\>|\cos\phi|$ in the linear expansion of the source
  (see \eqref{eq:linearU}) has to be replaced by
\begin{equation}
  \label{eq:coherence}
\sqrt{\ka^2+m^2}\>|\cos\phi|\to
\left|\sqrt{\ka^2+m^2}\,\cos\phi+\al p_{ax}\right|-\al|p_{ax}|\>,
\end{equation}
with $\al$ the Sudakov variable in the $ab$-dipole centre of mass in
the forward region $\al>\be$;
\item to take fully into account effects of non-inclusiveness of jet
  observables at two-loop order, we multiply the radiator by the Milan
  factor \cite{Milan,Milan2}
  \begin{equation}
    \label{eq:cM}
  \cM = \frac{3}{64}\frac{(128\pi+128\pi\ln 2-35\pi^2)C_A-5\pi^2
    n_f}{11C_A-2 n_f}\>,
\end{equation}
using $n_f=3$;
\item the NP correction is finally expressed in terms of the parameter
  \begin{equation}
    \label{eq:cp'}
\cp=2\cM\, c_{\Ko}\int dm\frac{\de\aef(m)}{\pi}\>, 
\quad c_{\Ko}=\frac{2}{\pi}\>.
\end{equation}
After merging PT and NP contributions to the observable in a
renormalon free manner, one has that the distribution is independent
of $\mu_I$ and one obtains
\begin{equation}
  \label{eq:cp}
 \cp \!\equiv  c_{\Ko}\,\cM \frac{4}{\pi^2}\mu_I
\left\{ \alpha_0(\mu_I)- \bar{\alpha}_s
  -\beta_0\frac{\bar{\alpha}_s^2}{2\pi}\left(\ln\frac{Q}{\mu_I} 
+\frac{K}{\be_0}+1\right) \right\},  
\end{equation}
where 
\begin{equation}
\bar{\alpha}_s\equiv \al_{\MSbar}(Q)\>,\quad
   K\equiv
  C_A\left(\frac{67}{18}-\frac{\pi^2}{6}\right)-\frac{5}{9}n_f \>,
  \quad \beta_0=\frac{11N_c}{3}-\frac{2n_f}{3}\>.
\end{equation}
The $K$ factor accounts for the mismatch between the $\MSbar$ and the
physical scheme \cite{CMW} and $\al_0(\mu_I)$ is the integral of the
running coupling over the infra-red region, see \eqref{eq:al0}.

The numerical coefficient $c_{\Ko}$ depends on our observable $\Ko$.
For instance, the shift for the $\tau=1\!-\!T$ distribution is
\begin{equation}
  \label{eq:thrust}
  \frac{d\sigma}{d\tau}(\tau)=
\frac{d\sigma^{\PT}}{d\tau}(\tau\!-\!\Delta_{\tau})\>,
\qquad   \Delta_{\tau}=C_F\,\frac{c_{\tau}\,\cp}{c_{\Ko}}\>,
\quad c_{\tau}=2\>,
\end{equation}
where $C_F$ enters due to the fact the $2$-jet system is made of
a quark-antiquark pair.

\end{itemize}
We recall that these prescriptions correspond to taking into account NP
corrections at two-loop order in the reconstruction of the
(dispersive) running coupling and in the non-inclusive nature of
the observable. 
We implement the rapidity cut by expressing $\eta_k$ the soft
gluon rapidity in the invariant form \eqref{eq:rapB}.

\subsection{Dipoles $12$ and $13$}
Consider first the contribution of the $1a$-dipole  ($a=2,3$).
We decompose the gluon momenta along $P_1$ and $P_a$ taken in their 
centre of mass 
system:
\begin{equation}
\label{eq:1aSud}
P_a^*=\frac{Q_{1a}}{2}(1,0,0,1)\>,\quad P_1^*=\frac{Q_{1a}}{2}(1,0,0,-1)\>,
\quad k^*=\al P_a^*+\be P_1^*+\ka\>.
\end{equation}
and we use the expression in \eqref{eq:coherence} in our linearized 
source, so that the recoil momentum of parton $p_a$ provides an 
effective rapidity cutoff for a gluon emitted
collinear to $P_a$.
In the region in which $k^*$ is emitted close to $P_1$, i.e. $\be\gg\al$,
 we can write gluon rapidity in terms of the Sudakov variables 
\eqref{eq:1aSud} as follows:
\begin{equation}
\label{eq:cut1a}
\eta_k=\frac12\ln\frac{(\xB P+q)k}{\xB (Pk)}
\simeq\ln \frac{Q\sqrt{\ka^2+m^2}}{\al Q_{1a}^2\ics}\>,
\end{equation}
so that the NP correction to the $1a$-dipole radiator is given by
\begin{equation}
\label{eq:der1a}
\begin{split}  
\de r_{1a}
&= \frac{\nu\cM}{\pi}\! \int \! dm^2\de\aef(m)\frac{-d}{dm^2}\!
\int \frac{d\ka^2}{\ka^2+m^2}\>I_{1a}\>,\\
I_{1a}&=\int_{-\pi}^{\pi }\frac{d\phi}{2\pi} 
\int_{0}^{1}\frac{d\al}{\al}
\left(\left|\sqrt{\ka^2\!+\!m^2}\cos\phi\!+\!\al p_{ax}\right|
\!-\!\al |p_{ax}|\right) \Theta\left(\eta_0+
\ln\frac{\al Q^2_{1a}\>\ics}{Q\sqrt{\ka^2+m^2}}\right).
\end{split}
\end{equation}
If we consider the two rapidity cutoffs discussed above we have that
the $\al$ integration is restricted to the region
\begin{equation}
\al_m<\al<\al_M\>,\qquad
\al_m\equiv
\frac{\sqrt{\ka^2+m^2}\,Q\,e^{-\eta_0}}{Q^2_{1a}\ics}\>,
\quad
\al_M \equiv \frac{\sqrt{\ka^2+m^2}}{|p_{ax}|}\>,
\end{equation}
giving 
\begin{equation}
  \label{eq:I1a}
\begin{split}
I_{1a}&=\int_{-\pi}^{\pi }\frac{d\phi}{2\pi} 
\int_{\al_m}^{\al_M}\frac{d\al}{\al}
\left(\left|\sqrt{\ka^2\!+\!m^2}\cos\phi\!+\!\al p_{ax}\right|
\!-\!\al |p_{ax}|\right) \\
&= \frac{2}{\pi}\sqrt{\ka^2+m^2}
\left(\eta_0+\ln\frac{Q_{1a}\ics}{Q}+
\ln\frac{Q_{1a}\zeta}{|p_{ax}|}\right),\qquad \zeta=2e^{-2}\>. 
\end{split}
\end{equation}
We thus obtain the NP correction to the $1a$-dipole radiator
\begin{equation}
\delta r_{1a}=\nu\cp\left(\eta_0+\ln\frac{Q_{1a}\ics}{Q}+
\ln\frac{Q_{1a}\zeta}{|p_{ax}|} \right)\>.
\end{equation}

\subsection{Dipole $23$}
Again we decompose the emitted gluon momentum along $P_2$ and $P_3$
\begin{equation}
\label{eq:23Sud}
P_2^*=\frac{Q_{23}}{2}(1,0,0,1)\>,\quad P_3^*=\frac{Q_{23}}{2}(1,0,0,-1)\>,
\quad k^*=\al P_2^*+\be P_3^*+\ka\>.
\end{equation}
In this case, if $k^*$ is emitted close to $P^*_2$ ($\al>\be$), its rapidity 
is cut by the recoil component $p_{2x}$, while it is $p_{3x}$ which provides
 an effective rapidity cutoff when $k^*$ is close to $P^*_3$ ($\al<\be$). 
Thus it is convenient to split the radiator into forward ($\al>\beta$) 
and backward ($\al<\beta$) regions. 
In the backward region one may relabel $\beta$ as $\al$. 
Performing the substitution in \eqref{eq:coherence} in the linearized 
source we obtain 
\begin{equation}
\label{eq:der23}
\begin{split}  
\de r_{23}
&= \frac{\nu\cM}{\pi}\! \int \! dm^2\de\aef(m)\frac{-d}{dm^2}\!
\int \frac{d\ka^2}{\ka^2+m^2}\>(I_{23}^2+I_{23}^3)\>,\\
I_{23}^a&=\int_{-\pi}^{\pi }\frac{d\phi}{2\pi} 
\int_{0}^{1}\frac{d\al}{\al}
\left(\left|\sqrt{\ka^2\!+\!m^2}\cos\phi\!+\!\al p_{ax}\right|
\!-\!\al |p_{ax}|\right) \Theta\left(
\al-\frac{\sqrt{\ka^2+m^2}}{Q_{23}}\right)\>,
\end{split}
\end{equation}
(for $a=2,3$),
so that the integration limits for $\al$ become
\begin{equation}
\al_m<\al<\al_M\>,\qquad
\al_m\equiv
\frac{\sqrt{\ka^2+m^2}}{Q_{23}}\>,
\quad
\al_M \equiv {\frac{\sqrt{\ka^2+m^2}}{|p_{ax}|}}\>,
\end{equation}
giving the following result
\begin{equation}
  \label{eq:I23a}
\begin{split}
I_{23}^a&=\int_{-\pi}^{\pi }\frac{d\phi}{2\pi} 
\int_{\al_m}^{\al_M}\frac{d\al}{\al}
\left(\left|\sqrt{\ka^2\!+\!m^2}\cos\phi\!+\!\al p_{ax}\right|
\!-\!\al |p_{ax}|\right) \\
&= \frac{2}{\pi}\sqrt{\ka^2+m^2}
\left(\ln\frac{Q_{23}\zeta}{|p_{ax}|}\right),\qquad \zeta=2\,e^{-2}\>. 
\end{split}
\end{equation}
Putting together the two pieces one obtains the NP correction to $r_{23}$
\begin{equation}
\de r_{23}=\nu\cp\left(\ln\frac{Q_{23}\zeta}{|p_{2x}|}
+\ln\frac{Q_{23}\zeta}{|p_{3x}|}\right)\>.
\end{equation}
In conclusion, assembling the contributions from the three dipoles,
one is left with the expression in \eqref{eq:RadNP}.

\section{Evaluating the NP shift \label{App:NPdist}}
Here we evaluate the NP shifts $\delta\Ko^{(\conf)}$ introduced in
\eqref{eq:ma}. Using the operator identity \eqref{eq:op-identity} 
we may write
\begin{equation}
\de\cA_{\conf}(\Ko)=\frac{-\cp}{\Ko}\frac{\frac{\partial}{\partial\ln\Ko}}
{\Gamma\left(1+\frac{\partial}{\partial\ln\Ko}\right)}\,f_\conf(\Ko^{-1})\>.
\end{equation}
Applying this differential operator yields
\begin{equation}
\begin{split}
&\de\cA_{\conf}(\Ko)=\frac{-\cp}{\Ko}\frac{1}{\Gamma(1+R')}
\prod_{a=2,3}\int_{-\infty}^{\infty}
\frac{d\be_a\,e^{-R_{1a}^{(\conf)}\left(\bKo^{-1}\sqrt{1+\be_a^2}\right)}}
{\pi(1+\be_a^2)}\Biggl[
C_1^{(\conf)}R'\ln\frac{\ics e^{\eta_0}Q_1^{(\conf)}}{Q}\\
&+\sum_{a=2,3}C_a^{(\conf)}\left\{R'\ln(\zeta\bKo^{-1}Q_a^{(\conf)})
+R'\psi(1+R_T')+R'\chi(\be_a)-1\right\}\Biggr]\,,
\end{split}
\end{equation}
with $\bKo=\Ko e^{-\gam_E}$, $R'=C_T r'(\Ko^{-1},Q)$ and 
$R_{1a}^{(\conf)}$ defined in \eqref{eq:R1a}.
The final term may now be integrated by parts to give
\begin{equation}
\begin{split}
&\de\cA_{\conf}(\Ko)=\frac{-\cp}{\Ko}\frac{R'}{\Gamma(1+R')}
\prod_{a=2,3}\int_{-\infty}^{\infty}
\frac{d\be_a\,e^{-R_{1a}^{(\conf)}\left(\bKo^{-1}\sqrt{1+\be_a^2}\right)}}
{\pi(1+\be_a^2)}\Biggl[
C_1^{(\conf)}\ln\frac{\ics e^{\eta_0}Q_1^{(\conf)}}{Q}\\
&+\sum_{a=2,3}C_a^{(\conf)}\left\{\ln(\zeta\bKo^{-1}Q_a^{(\conf)})
+\psi(1+R')+\half\ln(1+\be_a^2)+
\left(1\!-\!\frac{C_{1a}^{(\conf)}}{C_T}\right)
\be_a\tan^{-1}\be_a\right\}\Biggr]\>.
\end{split}
\end{equation}

For all but the final term in the brackets, we may expand the radiator
as in the PT calculation and perform the integrals over $\be_a$. The
final term is only slowly convergent and must be treated with extra care.
Thus
\begin{equation}
\de\cA_\conf(\Ko)=-\de\Ko^{(\conf)}\cdot\frac{R'}{\Ko}\cA_\conf^{\PT}(\Ko)
=-\de\Ko^{(\conf)}\cdot\partial_{\Ko}\cA_\conf^{\PT}
\end{equation}
with
\begin{equation}
\begin{split}
\frac{\de\Ko^{(\conf)}}{\cp} =
C_1^{(\conf)}\ln\frac{\ics e^{\eta_0}Q_1^{(\conf)}}{Q}
&+\sum_{a=2,3}C_a^{(\conf)}\left[\ln\frac{\zeta Q_a^{(\conf)}}{\bKo}+
\psi(1+C_T r')+\half\psi\left(1+\frac{C_{1a}^{(\conf)}r'}{2}\right)\right.\\
&\left.-\half\psi\left(\frac{1+C_{1a}^{(\conf)}r'}{2}\right)
+H_a^{(\conf)}(\bKo^{-1})\right]\>.
\end{split}
\end{equation}
The term called $H_a^{(\conf)}(\bKo^{-1})$ arises from this final
term. For the term proportional to $C_2^{(\conf)}$ we may expand the 
radiator in $\be_3$ and integrate over it, and vice versa, giving
\begin{equation}
H_a^{(\conf)}(\bKo^{-1}) = \left(1-\frac{C_{1a}^{(\conf)}}{C_T}\right)
\frac{e^{R_{1a}^{(\conf)}(\bKo^{-1})}}{\cF(C_{1a}^{(\conf)}r')}
\int_{-\infty}^{\infty}\frac{d\be_a\,e^{-R_{1a}^{(\conf)}
\left(\bKo^{-1}\sqrt{1+\be_a^2}\right)}}{\pi(1+\be_a^2)}\,
\be_a\tan^{-1}\be_a\>.
\end{equation}

The reason we may not simply expand the remaining radiator and integrate
over $\be_a$ is that doing so gives an unphysical result that
diverges in the limit $r'(\Ko^{-1})\to0$. This unphysical divergence is
regulated by the second derivative of the radiator. We write
\begin{equation}
\beta\tan^{-1}\beta = \frac{\pi}{2}|\beta|-\beta\tan^{-1}\frac{1}{\beta}\>.
\end{equation}
By making the substitution $\rho=\bKo^{-1}\sqrt{1+\be_a^2}$, the contribution
from the first term is written in terms of the function $E_a$, discussed in
the next appendix. The second contribution gives a fastly convergent
integral, so we can expand the radiator and integrate as before. We find
\begin{equation}
H_a^{(\conf)}(\bKo^{-1}) = \left(1-\frac{C_{1a}^{(\conf)}}{C_T}\right)
\frac{1}{\cF(C_{1a}^{(\conf)}r')}\left(E_a^{(\conf)}(\bKo^{-1})+
\frac{\cF(C_{1a}^{(\conf)}r')-1}{C_{1a}^{(\conf)}r'}\right)\>.
\end{equation}
where the $E_a$ functions are discussed below. Thus we recover the result
\eqref{eq:NPshift}.

\section{The $E_a$ functions \label{App:Ea}}
For the functions $E_a^{(\conf)}(\bKo^{-1})$ we obtain the following result: 
\begin{equation}
\label{eq:ea}
\begin{split}
&E_a^{(\conf)}(\bKo^{-1})\equiv\int_{\bKo^{-1}}^{\infty}\frac{d\rho}{\rho}\,
e^{-R_{1a}^{(\conf)}(\rho)+R_{1a}^{(\conf)}(\bKo^{-1})}
\left(\frac{\cP(\rho^{-1})}{\cP(\Ko)}\right)^{\frac12}\\
&=\sqrt{\frac{\pi}{2R_{1a}''}}\>N(t)
- \frac{R_{1a}'''}{3R_{1a}^{\prime\prime2}}\>X(t)+\cO{\sqrt{\as}}\>,
\quad t\equiv \frac{1}{\sqrt{2R_{1a}''}}
\left(R_{1a}'+\frac12 \frac{\partial\ln\cP(\Ko)}{\partial\ln\Ko}\right)\>,
\end{split}
\end{equation}
where  $R_{1a}',R_{1a}''$ and $R_{1a}'''$ are the first three logarithmic
derivatives of $R_{1a}^{(\conf)}(\rho)$ evaluated at $\rho=\bKo^{-1}$,
and the functions $N$ and $X$ are defined by
\begin{equation}\label{eq:N-X}
\begin{split}
N(t)\equiv\frac{2}{\sqrt{\pi}}\,\int_t^{\infty}dx\> e^{-x^2+t^2}\>,
\qquad 
X(t)\equiv
-\frac{\sqrt{\pi}}{8}\frac{d^3N(t)}{dt^3}=
\int_0^{\infty}dz\> z\>  e^{-z -2 t \sqrt{z}}\>.
\end{split}
\end{equation}
They have the following asymptotic behaviours
\begin{equation}
\label{eq:expansion}
\begin{split}
& t\gg1: \quad
N(t)= \frac{1}{\sqrt{\pi}t}\left(1-\frac{1}{2t^2}
+\frac{3}{t^4}+\dots\right), 
\quad X(t)= \frac{3}{4 t^4}+ \cO{t^{-6}}\>,\\
& t\ll1:\quad 
N(t)= 1-\frac{2t}{\sqrt{\pi}}+t^2-\frac{4t^3}{3\sqrt{\pi}}+\dots, 
\quad X(t)= 1-\frac{3\sqrt{\pi}}{2}t +4 t^2+\cO{t^3}\>.
\end{split}
\end{equation}

In the region $R_{1a}'/\sqrt{2R_{1a}''}\gg1$, using the expansions in 
\eqref{eq:expansion}, we get
\begin{equation}
  \label{eq:regione1a}
E_a^{(\conf)}(\bKo^{-1})=\frac{1}{R_{1a}'}
\left\{1+\cO{\frac{1}{R_{1a}^{\prime}}
\frac{\partial\ln\cP(\Ko)}{\partial\ln\Ko}}\right\}
=\frac{1}{C_{1a}^{(\conf)}r'}
\left\{1+\cO{\frac{1}{\ln(Q/\Ko)}}\right\}\>. 
\end{equation}
On the contrary, for $R_{1a}'/\sqrt{2R_{1a}''}\ll 1$, we
obtain, up to contributions $\cO{\sqrt{\as}\,}$, 
\begin{equation}
\label{eq:regione2a}
E_a^{(\conf)}(\bKo^{-1})=\sqrt{\frac{\pi}{2R_{1a}''}}
-\frac{R_{1a}'}{R_{1a}''}
-\frac{1}{2R_{1a}''}\frac{\partial\ln\cP(\Ko)}{\partial\ln\Ko}
-\frac{R_{1a}'''}{3R_{1a}^{\prime\prime2}}+\cdots\>.
\end{equation}
Starting from the definition \eqref{eq:R1a} we find
\begin{equation}
\begin{split}
R_{1a}' &= \frac{2}{\pi}\as(2\bKo)\left(
\half C_1^{(\conf)}\ln\frac{\zeta_1^{(\conf)}Q_1^{(\conf)}}{2\bKo}+
C_a\ln\frac{\zeta_a^{(\conf)}Q_a^{(\conf)}}{2\bKo}\right)\\
R_{1a}'' &= \frac{2}{\pi}C_{1a}^{(\conf)}\as(2\bKo)
+\frac{\be_0}{\pi^2} \as^2(2\bKo)\left(
\half C_1^{(\conf)}\ln\frac{\zeta_1^{(\conf)}Q_1^{(\conf)}}{2\bKo}+
C_a\ln\frac{\zeta_a^{(\conf)}Q_a^{(\conf)}}{2\bKo}\right)  \\
R_{1a}''' &= \frac{2\be_0}{\pi^2}C_{1a}^{(\conf)}\as^2+\cdots
\end{split}
\end{equation}
and therefore
\begin{equation}
\begin{split}
E_a^{(\conf)}(\bKo^{-1})= \frac{\pi}{2\sqrt{C_{1a}^{(\conf)}\as}} &-
\frac{1}{C_{1a}^{(\conf)}}\left(\!
\half C_1^{(\conf)}\ln\frac{\zeta_1^{(\conf)}Q_1^{(\conf)}}{2\bKo}\!+\!
C_a\ln\frac{\zeta_a^{(\conf)}Q_a^{(\conf)}}{2\bKo}
\right.\\
&\left.+\frac{\pi}{4\as}\frac{\partial\ln\cP(\Ko)}{\partial\ln\Ko}
+\frac{\be_0}{6}\right)
+\cO{\sqrt{\as}}\,. 
\end{split}
\end{equation}

\end{document}